\documentclass[%
 reprint,
 amsmath,amssymb,
 aps,
pra,
]{revtex4-2}

\usepackage{graphicx}
\usepackage{dcolumn}
\usepackage{bm}

\usepackage{braket}
\usepackage{physics}
\usepackage{hyperref}
\usepackage{url}
\usepackage{amsmath}
\usepackage{amssymb}
\usepackage{mathtools}
\usepackage{comment}
\usepackage{xcolor}

\usepackage{mathrsfs}
\usepackage{cancel}
\usepackage{soul}

\usepackage[linesnumbered,titlenumbered,ruled,vlined,resetcount]{algorithm2e}
    
\begin{document}

\preprint{APS/123-QED}

\title{Mitigating shot noise in local overlapping quantum tomography with semidefinite programming}

\author{Zherui Jerry Wang$^{1,2,4}$}%
\email{zherui@lorentz.leidenuniv.nl}
\author{David Dechant$^{1,2}$}%
\author{Yash J. Patel$^{1,3}$}%
\author{Jordi Tura$^{1,2}$}%
\email{tura@lorentz.leidenuniv.nl}
\affiliation{$^1$ $\langle aQa^L \rangle$ Applied Quantum Algorithms, Universiteit Leiden, the Netherlands}%
\affiliation{$^2$Instituut-Lorentz, Universiteit Leiden, P.O. Box 9506, 2300 RA Leiden, The Netherlands}%
\affiliation{$^3$LIACS, Universiteit Leiden, Niels Bohrweg 1, 2333 CA Leiden, Netherlands}%
\affiliation{$^4$QuTech, Delft University of Technology, P.O. Box 5046, Delft 2600 GA, Netherlands}

\date{\today}

\begin{abstract}
Reduced density matrices (RDMs) are fundamental in quantum information processing, allowing the computation of local observables, such as energy and correlation functions, without the exponential complexity of fully characterizing quantum states. 
In the context of near-term quantum computing, RDMs provide sufficient information to effectively design variational quantum algorithms.
However, their experimental estimation is challenging, as it involves preparing and measuring quantum states in multiple bases--a resource-intensive process susceptible to producing non-physical RDMs due to shot noise from limited measurements.
To address this, we propose a method to mitigate shot noise by re-enforcing certain physicality constraints on RDMs. 
While verifying RDM compatibility with a global state is quantum Merlin-Arthur complete, we relax this condition by enforcing compatibility constraints up to a certain level using a polynomial-size semidefinite program to reconstruct overlapping RDMs from simulated data.
Our approach yields, on average, tighter 
 bounds for the same number of measurements compared to tomography without compatibility constraints. 
We demonstrate the versatility and efficacy of our method by integrating it into an algorithmic cooling procedure to prepare low-energy states of local Hamiltonians. 
Simulations on frustrated Hamiltonians reveal notable improvements in accuracy and resource efficiency, highlighting the potential of our approach for practical applications in near-term quantum computing.
\end{abstract}

\maketitle

\section{Introduction}\label{sec:introduction}

Accurately characterizing quantum states is crucial for quantum computing, and can be achieved by quantum state tomography methods \cite{nielsen2010quantum, lvovsky2009continuous, cramer2010efficient}. It has applications ranging from certifying quantum devices \cite{PRXQuantum.2.010201} to the design and execution of variational quantum algorithms \cite{cerezo2021variational}.
However, the full characterization of a quantum state becomes impractical for larger systems. 
This is due to an exponential growth in the number of required measurement settings, which correspond to distinct Pauli strings.
To address this challenge, several alternative strategies have been proposed, such as classical shadow tomography \cite{Huang_2020}, which significantly reduce the number of required measurement settings.
However, even with these techniques, each measurement setting must be typically repeated for a number of shots, $N_{meas}$, to achieve the desired accuracy, with the statistical uncertainty scaling as $\mathcal{O}(1/\sqrt{N_{meas}})$.
Since the measurements collapse the quantum state, the state must be re-prepared for each shot.

Many applications, such as ground state optimization of local Hamiltonians \cite{liu2021efficient,araujo2022local}, rely on the tomography of $k$-qubit reduced density matrices (RDMs), which describe a subsystem of the full quantum state \cite{davidson2012reduced}.
The estimation of all $k$-qubit RDMs of an $n$-qubit system, can be achieved by using $\binom{n}{k} \cdot 3^k$ distinct measurement settings \cite{linden2002almost, linden2002parts, cotler2020quantum}.
However, even with exact estimates of RDMs, the full characterization of a general quantum state would entail resolving the quantum marginal problem \cite{klyachko2006quantum}, a task known to be quantum Merlin-Arthur (QMA) complete \cite{broadbent2022qma,kamminga2024complexitypurestateconsistencylocal}.
These extensive measurement requirements pose a bottleneck for applying variational quantum algorithms in scenarios requiring high-precision results, such as in fields like computational chemistry, where a precision of
$10^{-3}$ Hartree \cite{mccaskey2019quantum} (chemical accuracy) is often the target.

Recent strategies like overlapping tomography \cite{cotler2020quantum, araujo2022local, hansenne2024optimal} have emerged to minimize the total number of measurement settings while maintaining high precision. 
These methods achieve this by parallelizing measurements on non-overlapping subsystems and efficiently organizing information from measurements on overlapping subsystems.
However, current methods, while promising, often overlook higher-order correlations and compatibility constraints among RDMs, potentially limiting their accuracy.

In this work, we propose a hierarchy of data-driven semidefinite programs (SDPs) to estimate a set of overlapping reduced density matrices (RDMs) from quantum measurements.
Our approach focuses specifically on random Pauli string measurements of $n$-qubit states with fixed locality.
We leverage the inherent higher-order correlations present in quantum state measurement data--information that would be lost when estimating overlapping RDMs independently.
This enables us to tighten the uncertainty intervals of the RDM estimates for a given number of measurement shots, with particular advantages in low-shot regimes.
Such SDP relaxations are built on re-imposing partial compatibility with the quantum marginal problem \cite{skrzypczyk2023semidefinite, kamminga2024complexitypurestateconsistencylocal,aloy2021quantum}. 
We further constrain each RDM to satisfy physical requirements of a valid density matrix, namely unit trace and positive semidefiniteness.
This comprehensive approach yields two key benefits: it resolves the compatibility issues between overlapping RDMs that arise in linear inversion \cite{schwemmer2015systematic}, while simultaneously enhancing the global consistency of the entire set of RDMs.

Our method demonstrates superior performance across two numerical benchmarks.
In the first evaluation, we assess the ability of our method to estimate ground-state RDMs and energies of the one-dimensional (1D) \(XY\) model with open boundary conditions by using random Pauli string measurements.
Under the same measurement budget, our approach achieves more accurate energy estimates compared to conventional tomography methods.
We further validate our technique through application to algorithmic cooling \cite{boykin2002algorithmic, polla2021quantum, grimsley2019adaptive}, a practical use case in near-term quantum computing where RDMs inform quantum circuit design. 
In this application domain, our method again demonstrates measurable advantages over traditional approaches that rely on independent RDM reconstructions.

This paper is organized as follows: Section \ref{sec:background} introduces quantum tomography, semidefinite programming, and related literature. 
Section \ref{sec:method} presents the proposed SDP-based reconstruction method and numerical results. 
In Section \ref{sec:app}, we apply our method to algorithmic cooling. 
Finally, Section \ref{sec:outlook} provides conclusions and future directions.

\section{Background and preliminaries}
\label{sec:background}
\subsection{Notations}
We consider an $n$-qubit system acting on the composite Hilbert space ${\cal H} = ({\mathbb C}^2)^{\otimes n}$.
For our mathematical representation, we outline the relevant definitions and notations below.
Let $\sigma_{0,1,2,3}$ denote, respectively, the identity matrix $\mathbb{I}$ and the Pauli $X$, $Y$, $Z$ matrices. 
For an $n$-qubit system, the identity matrix is denoted by $\mathbb{I}_n$.
A vector $\mathbf{i}=(i_1,i_2,...,i_n) \in \{0,1,2,3\}^{n} \coloneqq \mathcal{I}_n$ is used to represent a specific Pauli basis $\sigma_{\mathbf{i}}$, which is a shorthand for the Pauli string $\sigma_{i_1}\otimes\sigma_{i_2}\otimes ...\otimes \sigma_{i_n}$,
in $\{\mathbb{I},X,Y,Z\}^n$. 
For simplicity, we use this notation throughout the text. 
Note that Pauli strings form an orthogonal basis of the $\mathbb{R}$-vector space of Hermitian matrices.
We denote the set $\{1,...,m\}$ by $[m]$.
For a Hermitian matrix $A$, the notation $A \succeq 0$ indicates that $A$ is positive semidefinite, i.e., all its eigenvalues are non-negative.
We denote by $\rho_{AB}$ the density matrix that describes a quantum state in the Hilbert space $\mathcal{H}_{AB} = \mathcal{H}_{A} \otimes \mathcal{H}_{B}$, where $A = \{a_1, \ldots, a_{|A|}\}$ and $B = \{b_1, \ldots, b_{|B|}\}$ label distinct sets of qubits in the respective subsystems.
The partial trace over subsystem $A$ is defined as:
\begin{align}
    \Tr_{A}\left[\rho_{AB}\right]
    \coloneqq
    \sum_{k}(\bra{k}\otimes \mathbb{I}_B)\rho_{AB}(\ket{k}\otimes \mathbb{I}_B),
\end{align}
where $\{\ket{k}\}$ forms an orthonormal basis of the Hilbert space $\mathcal{H}_{A}$.

\subsection{Quantum State Tomography}

Quantum state tomography (QST) \cite{nielsen2010quantum, lvovsky2009continuous, cramer2010efficient}
is a cornerstone of quantum information science, enabling the reconstruction of quantum states through systematic measurements on an ensemble of identical quantum states.
For an $n$-qubit quantum state, its density matrix, represented by $\rho$, can be fully characterized by the following relation:
\begin{align}
    \rho =& \frac{1}{2^n} \left(\sum_{ \mathbf{i} \in \mathcal{I}_n }C_{\mathbf{i}} 
    \sigma_{\mathbf{i}}\right), \\
    C_{\mathbf{i}} =& \Tr[\rho \sigma_{\mathbf{i}}].
\end{align}
Here, $C_{\mathbf{i}}$ constitutes an element of the corresponding $n$-qubit Bloch vector, with
$C_{\mathbf{0}} = 1$.
In practical settings, $C_{\mathbf{i}}$ is not directly accessible, instead an estimate $\hat{C}_{\mathbf{i}}$ is computed by averaging over a finite number of measurements $N_{\text{meas},\mathbf{i}}$ performed in the basis $\sigma_{\mathbf{i}}$:
\begin{align}
\hat{C}_{\mathbf{i}} = \frac{1}{N_{\text{meas},\mathbf{i}}} \sum_{k=1}^{N_{\text{meas},\mathbf{i}}} m_{\mathbf{i}}^k,
\end{align}
where $m_{\mathbf{i}}^k\in \{1, -1\}$ represents the $k$-th measurement outcome. 
In measuring a particular Pauli string $\sigma_{\mathbf{i}}$, one obtains a binary string $s_{\sigma_{\mathbf{i}}}=\{0,1\}^n$, from which the outcome
\begin{align}
    m_{\mathbf{i}}^k=(-1)^{|s_{\sigma_{\mathbf{i}}}|},
\end{align}
is calculated, with $|s_{\sigma_{\mathbf{i}}}|$ denoting the string Hamming weight, defined as the total count of $1$ s in $s_{\sigma_{\mathbf{i}}}$.
We denote the reconstructed state as $\hat{\rho}$, characterized by the $4^n-1$ different $\{\hat{C}_{\mathbf{i}}\}_{\mathbf{i}\in \mathcal{I}_n\setminus \{\mathbf{0}\}}$. 
Due to redundancies \cite{altepeter20044}, it is sufficient to simulate measurements in $3^n$ different bases, each corresponding to a specific Pauli string $\sigma_{\mathbf{i}}$ with $\mathbf{i} \in \{1,2,3\}^n$,
to estimate all $\hat{C}_{\mathbf{i}}$.
 In essence, if a string contains a $0$, then the corresponding qubits are not measured, which is equivalent to measuring the RDM of the complementary subsystem.
The measurement data collected from the other measurements are sufficient for reconstructing this RDM.
The reconstructed state $\hat{\rho}$ takes the form:
\begin{align}
    \hat{\rho} = \frac{1}{2^n} \left( \mathbb{I}_n + \sum_{ \mathbf{i} \in \mathcal{I}_n \setminus \{\mathbf{0}\}} \frac{1}{N_{\text{meas},\mathbf{i}}}\sum_{k=1}^{N_{\text{meas},\mathbf{i}}} m_{\mathbf{i}}^k 
    \sigma_{\mathbf{i}}\right).
    \label{eq:rho_hat_tomo}
\end{align}
According to the Chernoff-Hoeffding bound, achieving an additive error $\epsilon$ in each $\hat{C}_{\mathbf{i}}$ with respect to the true values $C_{\mathbf{i}}$ requires on the order of 
\mbox{$N_{\text{meas}}\approx 4\ln (2) n/\epsilon^2$} shots~\cite{cotler2020quantum} for a constant failure probability.
If each $\hat{C}_{\mathbf{i}}$ is estimated with $N_{\text{meas}}$ shots, then $3^n\,N_{\text{meas}}$ total measurements are required, because there are $3^n$ non-identity Pauli strings, rendering QST infeasible for larger $n$. 
For instance, a $50$-qubit system would demand $3^{50} \approx 7.18 \times 10^{23}$ measurement settings, and even with the Google Sycamore chip's $4$-$\mu s$ \cite{sycamoredatasheet} readout time, performing just one shot per setting would take about $10^{10}$ years. 

To ease this exponential scaling  in the number of qubits $n$, various methods have been proposed. 
Most notably, classical shadow tomography~\cite{Huang_2020} provides a way to extract certain few-body observables with a number of measurements that grows only logarithmically in the number of those observables.
Other techniques, such as neural-network-based tomography~\cite{torlai2018neural}, rely on sampling protocols and often presume certain structural features of the underlying system.

In Section~\ref{subsection: overlapping tomography}, we will present the overlapping tomography method of \cite{cotler2020quantum}, which can be applied for quantum state tomography of RDMs.

\subsection{Semidefinite Programming} 

A semidefinite-programming (SDP) problem involves optimizing a linear function subject to constraints expressed as linear matrix equalities and inequalities. 
The feasible regions of an SDP are known as spectrahedra, and efficient algorithms such as interior point methods~\cite{karmarkar1984new, jiang2020faster} can be used to solve them.
In its standard primal form, an SDP can be written as:
\begin{align}
\min \quad \langle C, X \rangle& \label{eq:sdp general} \\
\textrm{s.t.} \quad  \langle A_i, X \rangle& = b_i, \quad i = 1,2,...,m \notag, \\
    X \succeq 0 \notag.
\end{align}
where the cost matrix $C$, the decision matrix $X$, and the constraint matrices $A_i$ are Hermitian, $b$ is a real
vector, and $\langle \cdot , \cdot \rangle$ denotes the Hilbert-Schmidt inner product, defined as $\langle X, Y \rangle \coloneqq \Tr[X^{\dagger}Y ]$, with $X^{\dagger}$ being the Hermitian conjugate of $X$.
SDPs are particularly well-suited for quantum information~\cite{skrzypczyk2023semidefinite}, as density matrices are positive semidefinite. 
For instance, the quantum marginal problem seeks to determine whether a global state on a full system exists that is consistent with specified marginal states on subsystems, represented as reduced density matrices (RDMs)~\cite{skrzypczyk2023semidefinite, kamminga2024complexitypurestateconsistencylocal, aloy2021quantum}. 
This problem can also be formulated as an SDP, as illustrated by the following example.

Assume we have perfect knowledge of the RDMs \(\rho_{23}\), \(\rho_{13}\), and \(\rho_{12}\), acting on subsystems \(\mathcal{H}_{23}\), \(\mathcal{H}_{13}\), and \(\mathcal{H}_{12}\), respectively, and the goal is to find a global state \(\rho_{123}\) on \(\mathcal{H}_{123}\) that is consistent with these RDMs:
\begin{align}
\label{eq:sdp feasibility}
\mathrm{min}_{\rho_{123}} & 0\\
\textrm{s.t.} 
\quad & \rho_{123} \succeq 0,  \quad \Tr\left[\rho_{123}\right]=1, \notag \\
\quad & \Tr_{1}\left[\rho_{123}\right] = \rho_{23}, \notag\\
\quad & \Tr_{2}\left[\rho_{123}\right] = \rho_{13}, \notag\\
\quad & \Tr_{3}\left[\rho_{123}\right] = \rho_{12}. \notag
\end{align}
Here, the minimization of the constant \(0\) is purely a formalism to match the standard SDP form in Eq.~\eqref{eq:sdp general}. If the constraints cannot be satisfied, the SDP is deemed infeasible. Despite the apparent simplicity of compatibility SDPs like Eq.~\eqref{eq:sdp feasibility}, the general problem of determining whether a state \(\rho_{123\ldots n}\) exists is QMA-complete~\cite{broadbent2022qma, kamminga2024complexitypurestateconsistencylocal}, making it computationally intractable even for quantum computers.

To address this complexity, we relax the problem in our method by using quantum state tomography estimates and accounting for their error ranges. Additionally, reinforcement learning has been shown to improve the selection of constraints that enforce compatibility with the quantum marginal problem in reconstructed solutions~\cite{requena2023certificates}.

\subsection{Overlapping tomography and related literature}
\label{subsection: overlapping tomography}

The quantum state tomography of a full quantum state requires a number of measurements that scales as $3^n$, where $n$ is the number of qubits \cite{altepeter20044}. 
However, if the goal is to estimate only the $\binom{n}{k}$ different $k$-qubit reduced density matrices (RDMs), each RDM requires only $3^k$ different measurement settings. 
The total number of measurements, however, includes an additional overhead due to the number of shots needed to estimate each measurement setting.

Several strategies have been proposed to optimize the selection of measurement settings and minimize the total number of measurements required. 
In \cite{cotler2020quantum}, a method called \textit{quantum overlapping tomography} was introduced for estimating $k$-qubit RDMs.
The key idea is to measure each qubit individually in a chosen basis, allowing non-overlapping $k$-qubit subsystems to be measured in parallel. 
Overlapping $k$-qubit subsystems are then reconstructed through classical post-processing. 
This approach reduces the required number of measurement settings to at most $\exp^{\mathcal{O}(k)}\ln^2(n)$, which is even more efficient than shadow tomography \cite{Huang_2020}, where the scaling is $\mathcal{O}(n\text{polyln}(n))$. 
The measurement selection is determined by a family of hash functions $(n, k)$, which partition the system into $k$ subsystems, with all qubits in each subsystem measured in the same basis. All $k$-qubit RDMs can then be reconstructed from this carefully chosen dataset. 
In a similar way, partitioning-based methods \cite{bonet2020nearly} lead to a scaling of the number of measurement settings needed as $\mathcal{O}\left(3^k\ln ^{k-1}n\right)$.

However, in many practical scenarios, such as systems with $k$-local Hamiltonians and nearest-neighbor interactions, only $k$-geometrically-local RDMs, those involving neighboring qubits, are required. 
This further reduces the number of measurement settings needed to a constant scaling with respect to the number of qubits $n$ \cite{araujo2022local}. 
Recent advances have shown, both theoretically and experimentally, that leveraging graph theory can optimize this method, allowing even $k$-qubit RDMs to be estimated with a number of measurement settings that remains constant regardless of the number of qubits $n$ \cite{hansenne2024optimal}.

Other approaches have focused on reducing the statistical uncertainty due to shot noise in various contexts, such as quantum state tomography \cite{de2023comparison}, ground-state estimation \cite{westerheim2023dual}, and calculations of fidelity and von Neumann entropy \cite{zambrano2023certification}. 
SDP has also been applied to quantum marginal problems for tasks like ground-state estimation \cite{wang2024certifying}, incorporating tensor network methods \cite{kull2024lower}, or imposing entropy constraints on RDMs \cite{fawzi2023entropy}.

Our method builds on the principle of measuring the entire quantum system via product measurements on single qubits and performing post-processing to estimate $k$-local RDMs. 
However, rather than focusing on optimizing the selection of measurement settings, we introduce physicality and consistency constraints on the estimated RDMs. 
This approach explicitly accounts for the shot noise arising from a limited number of shots per measurement setting, providing a clearer benchmark for practical applications.

\section{SDP-assisted overlapping Tomography \label{sec:method}}

In this section, we introduce the SDP-assisted overlapping tomography framework for estimating local RDMs of $n$-qubit states described by local Hamiltonians. 
We begin by presenting the hypergraph representation of local Hamiltonians in Section~\ref{sec:local_h}. 
In Section~\ref{sec:methodology}, we detail the core methodology of SDP-assisted overlapping tomography. 
Finally, to illustrate its effectiveness, we apply the approach to the ground states of the 1D \(XY\) model and discuss the numerical results in Section~\ref{sec:sdp_result}.

\subsection{Local Hamiltonian \label{sec:local_h}}
\begin{figure}[htb!]
    \centering
    \includegraphics[width=0.98\columnwidth]{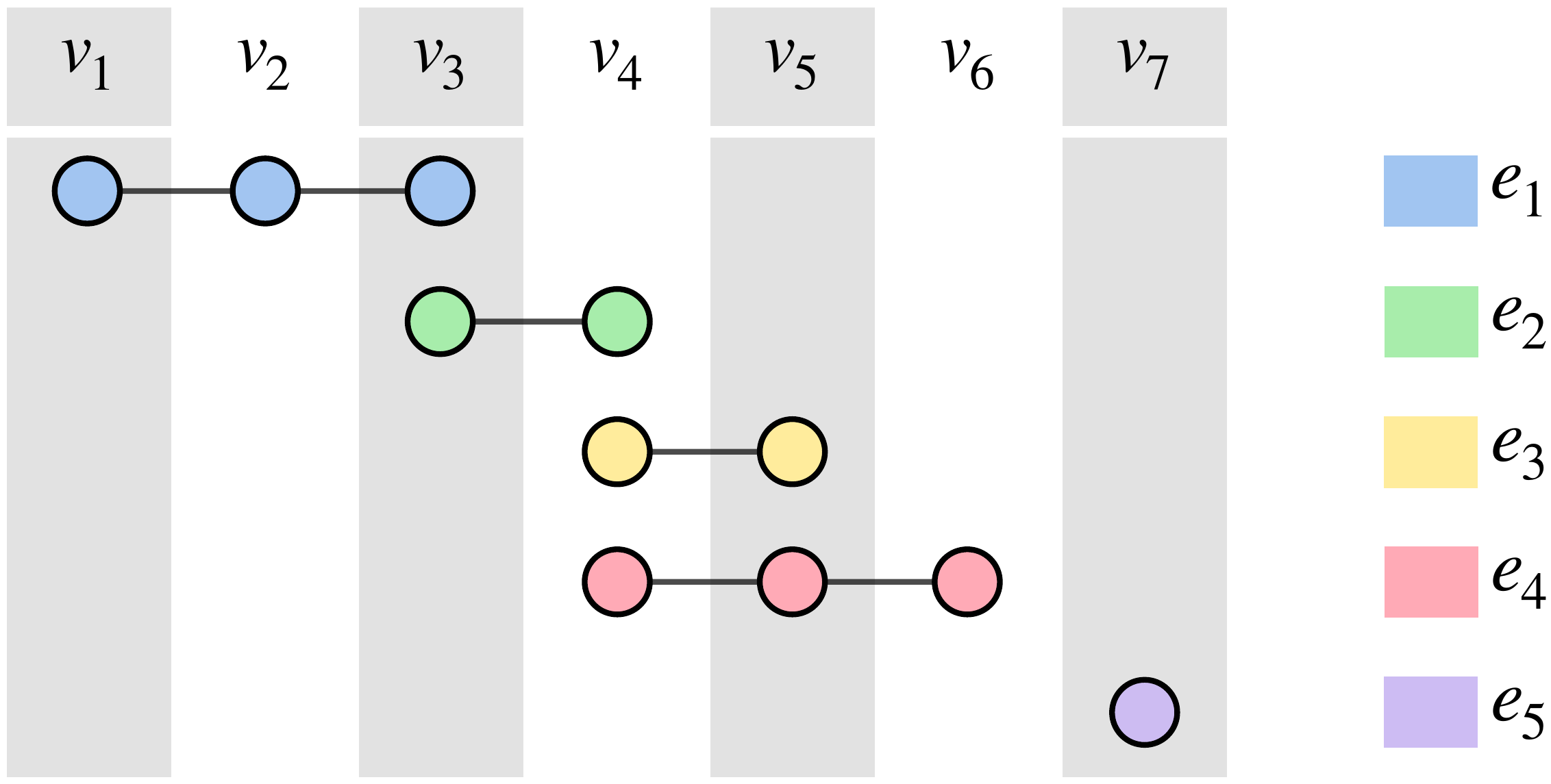}
    \caption{\textbf{Hypergraph representation of the interaction structure of a Hamiltonian $H=\sum_{j=1}^{m} H_j$.}
    The vertices $V_G=\{v_i\}_{i=1}^{n}$ correspond to $n=7$ qubits. 
    Each hyperedge $e_j \in E_G=\{e_j\}_{j=1}^{m}$ denotes a local Hamiltonian term $H_j$. 
    In this example, $E_G$ contains the sets $\{v_1,v_2,v_3\}$, $\{v_3,v_4\}$, $\{v_4,v_5\}$, $\{v_4,v_5,v_6\}$, $\{v_7\}$, hence $m=5$.
    Hyperedges are shown as horizontal lines linking the relevant vertices, and the corresponding local Hamiltonian term is indicated in the legend on the right.
    }
    \label{fig:hypergraph}
\end{figure}

The Hamiltonian of an $n$-qubit $k$-local system can be represented by a hypergraph $G = (V_G,E_G)$, where each vertex $v_i \in V_G$ corresponds to a qubit, and each hyperedge $e_j \in E_G$ (connecting up to $k$ vertices) represents a local Hamiltonian term.
Concretely, the Hamiltonian takes the form
\begin{align}
    H=\sum_{j=1}^{m} H_j. \label{eq:H_local}
\end{align}
where $m$ is the number of local subsystems, growing polynomially with $n$, and each $H_j$ acts non-trivially on at most $k$ qubits (i.e., it is $k$ local).
Fig. \ref{fig:hypergraph} illustrates the hypergraph representation of a \(k\)-local Hamiltonian in a many-body system.

The hypergraph representation not only provides a visualization of the Hamiltonian structure but also gives a framework for reconstructing the relevant RDMs.
Once measurement results are obtained from an ensemble of identical $n$-qubit states, one can reconstruct a collection of RDM estimates $\hat{\rho}_j$ for each local subsystem. 
This involves applying Eq.~\eqref{eq:rho_hat_tomo} to each of the $m$ subsystems:
\begin{align}
\left\{
\hat{\rho}_j \coloneqq 
\frac{1}{2^{|e_j|}} 
\left(
\mathbb{I}_{|e_j|} + \sum_{\mathbf{i} \in \mathcal{I}_{|e_j|}\setminus\{\mathbf{0}\}}\hat{C}^j_{\mathbf{i}} \sigma_{\mathbf{i}}
\right)
\right\}
_{j=1}^{m}.
\end{align}

Here, $\hat{C}^j_{\mathbf{i}}$ is the estimator of $C^j_{\mathbf{i}} = \Tr\left[\Tr_{V_G\setminus e_j}[\rho]\sigma_{\mathbf{i}}\right]=\Tr[\rho_j\sigma_{\mathbf{i}}]$, where $\rho$ is the true $n$-qubit global state and $\rho_j$ is the true RDM of the $j$-th subsystem.
Note that, each hyperedge $e_j = \{v_{j_1}, v_{j_2}, \dots\}$ corresponds to the qubits on which $H_j$ acts. 
These local RDMs are of particular interest because they are sufficient to estimate the energy expectation values of the local Hamiltonian:
\begin{align}
    \hat{E} = \sum_{j=1}^{m}\Tr[ \hat{\rho}_j H_j ]. \label{eq:exp_local_rdm}
\end{align}

However, the reconstructed local RDMs $\{\hat{\rho}_j\}_{j=1}^m$ can be non-physical in practice; e.g., they may exhibit negative eigenvalues or fail to satisfy mutual compatibility (i.e., overlapping RDMs might not be consistent). 
While mutual compatibility can only be enforced approximately in real scenarios~\cite{requena2023certificates}, imposing physicality constraints remains crucial for ensuring reliable energy estimates and other derived properties.

\subsection{Methodology\label{sec:methodology}}
In this section, we present the SDP-assisted overlapping tomography framework. 
Building on standard quantum tomography results, we formulate semidefinite programs (SDPs) 
that incorporate overlapping-compatibility (OC) and enhanced-compatibility (EC) constraints. 
These constraints aim to ensure the validity and consistency of the reconstructed states, allowing us to minimize or maximize the energy expectation within the feasible set. 
In essence, we determine the minimum and maximum energies compatible with the simulated data. 
By solving these SDP problems, we obtain a set of overlapping local RDMs that are more physically valid and mutually consistent.

\subsubsection{Semidefinite relaxations}

Relaxations of polynomial optimization problems based on semidefinite constraints play a central role in our method, in a similar spirit as Refs.~\cite{wang2024certifying, barthel2012solving}.
In this section, we describe the construction of these constraints.

Consider a local Hamiltonian associated with hypergraph $G=(V_G,E_G)$.
For each hyperedge $e_j \in E_G$, we introduce the decision variables for the SDP problem as
\begin{align}
\left\{
\Tilde{\rho}_j \coloneqq 
\frac{1}{2^{|e_j|}} 
\left(
\mathbb{I}_{|e_j|} + \sum_{\mathbf{i} \in \mathcal{I}_{|e_j|}\setminus\{\mathbf{0}\}}\Tilde{C}^j_{\mathbf{i}} \sigma_{\mathbf{i}}
\right)
\right\}
_{j=1}^{m},
\label{eq:decision_variable}
\end{align}
where
\begin{align} \label{eq:C_interval}
\Tilde{C}^j_{\mathbf{i}} \in \left[\hat{C}^j_{\mathbf{i}}-\epsilon^j_\mathbf{i},  \hat{C}^j_{\mathbf{i}}+\epsilon^j_\mathbf{i} \right], \quad \forall j \in [m],\ \forall \mathbf{i} \in \mathcal{I}_{|e_j|} \setminus \{\mathbf{0}\}.
\end{align}
Here, index $j$ labels the local subsystem associated with the hyperedge $e_j$, and $\epsilon^j_{\mathbf{i}}$ is a relaxation variable associated to $j$-th local subsystem. 
It is proportional to the variance of $\hat{C}^j_{\mathbf{i}}$, scaled by a coefficient $\alpha$, i.e., 
$\epsilon^j_\mathbf{i} = \alpha\,\mathrm{Var}(\hat{C}^j_{\mathbf{i}})$. 
The value of $\alpha$ relates to the probability that the unbiased estimator $\hat{C}^j_{\mathbf{i}}$ (of $\mathrm{Tr}[\rho_j \sigma_{\mathbf{i}}]$) lies within the chosen confidence interval, following Chernoff bound arguments. 
Throughout, tildes (\(\Tilde{\cdot}\)) denote SDP decision variables, while hats (\(\hat{\cdot}\)) denote results obtained directly from the raw data
during quantum state tomography.

Moreover, we introduce the semidefinite constraints 
\begin{align}\label{eq:sdp constraints}
    \Tilde{\rho}_j \succeq 0, \quad \forall j \in [m],
\end{align}
and the overlapping-compatibility (OC) and enhanced-compatibility (EC) constraints 
\begin{align}
    \mathscr{R}\coloneqq &\cup_{j,j' \in [m]}\mathcal{R}\left( \Tilde{\rho}_j, \Tilde{\rho}_{j'} \right), \\
    \mathscr{G}\coloneqq&\cup_{j,j' \in [m]
    }\mathcal{G}\left( \Tilde{\rho}_j, \Tilde{\rho}_{j'}\right),
\end{align}
respectively.
Specifically, $\mathcal{R}\bigl( \Tilde{\rho}_j, \Tilde{\rho}_{j'} \bigr)$ is the set of matrix equality constraints ensuring that the local RDMs are consistent on the intersection of their respective supports:
\begin{align}\label{eq:sdp_physical_constraints}
\mathcal{R}\left( \Tilde{\rho}_j, \Tilde{\rho}_{j'} \right) \coloneqq \{
\Tr_{e_j\setminus e_{j'}}\left[ \Tilde{\rho}_j \right]-\Tr_{e_{j'}\setminus e_{j}}\left[ \Tilde{\rho}_{j'} \right] = 0\},
\end{align}
while $\mathcal{G}\bigl( \Tilde{\rho}_j, \Tilde{\rho}_{j'} \bigr)$ is the set of constraints that require the existence of a larger RDM \(\Tilde{\rho}\) whose marginals agree on the variables indexed by $j$ and $j'$:
\begin{align}\label{eq:sdp_global_constraints}
\mathcal{G}\left( \Tilde{\rho}_j, \Tilde{\rho}_{j'}\right) \coloneqq \
\{
\Tilde{\rho} \succeq 0,  & \Tr_{e_j}\left[{\Tilde{\rho}}\right] = \Tilde{\rho}_{j'}, \Tr_{e_{j'}}\left[{\Tilde{\rho}}\right] = \Tilde{\rho}_{j}
\}.
\end{align}

Note that the dimension of $\Tilde{\rho}$ can freely be chosen and that the definition of $\mathcal{G}$ in Eq.~\eqref{eq:sdp_global_constraints} is just one possible choice; in practice, there is a hierarchy of possible relaxations forming a partially ordered set. 
The optimal choice of \(\mathcal{G}\) for a given computational budget is non-trivial, and one can achieve varying degrees of performance by judiciously selecting or refining these constraints~\cite{requena2023certificates, liu2007quantum}.

\subsubsection{SDP problem formulation}

The relaxations defined in the previous sub-section allow us to specify a feasible set of local RDMs.
In this work, we focus on finding the minimum and maximum energies consistent with the simulated data used to estimate the tomographic local RDMs \(\{\hat{\rho}_j\}\). Specifically, we consider the following optimization problem for determining the optimal state that minimizes (or maximizes) the energy under these constraints:
\begin{align}\label{eq:SDP_problem}
&\min(\max)_{\left\{ \Tilde{\rho}_j \right\}}
\quad \sum_j \Tr\left[\Tilde{\rho}_j H_j\right] \\
\textrm{s.t.} 
\ & \Tilde{C}^j_{\mathbf{i}} \in \left[\hat{C}^j_{\mathbf{i}}-\epsilon^j_\mathbf{i},  \hat{C}^j_{\mathbf{i}}+\epsilon^j_\mathbf{i} \right], \ \forall j\in [m] \ \text{and} \ \forall \mathbf{i} \in \mathcal{I}_{|e_j|} \setminus \{\mathbf{0}\},\notag \\
\quad & \Tilde{\rho}_j \succeq 0, \ \ \forall j \in [m], \notag\\
&{\mathscr R}\cup 
{\mathscr G}.\notag
\end{align}

Note that Eq.~\eqref{eq:SDP_problem} does not yet incorporate all the SDP constraints derived from the quantum marginal problem needed to solve the above problem, as it only includes the initial-order EC constraints. 
Obtaining the exact solution requires higher-order EC constraints, but adding these leads to exponential growth in computational complexity~\cite{liu2007quantum, requena2023certificates}. 
In large many-body systems, our formulation effectively discards higher-order constraints as \(n\) grows but \(k\) remains fixed, which loosens the energy bounds. Depending on the specific system, one may selectively include higher-level constraints that are most relevant for the problem at hand, here being the estimation of the minimum and maximum energy compatible with the measurement data~\cite{kull2024lower}.
  Such strategies (especially the reinforcement-learning-based \cite{requena2023certificates} and the renormalization-based \cite{kull2024lower} ones) can naturally be incorporated in our framework, thereby allowing for even tighter estimates, but this would impair the fairness of the benchmarking of our method. Hence, in the rest of the paper we proceed with the straightforward approach.

Because shot noise affects the data, setting \(\epsilon^j_{\mathbf{i}}=0\) (i.e., assuming no error) typically makes the SDP infeasible. 
Allowing a tolerance \(\epsilon^j_{\mathbf{i}}\) around the estimated quantities provides a search
region for solutions. 
As more data are collected, the \(\epsilon\) values shrink, thereby reducing the volume of the feasible set. 
Since the true state arises from a valid quantum system (satisfying the quantum marginal problem at all hierarchy levels), the resulting energy bounds become more accurate with increased data.

\subsubsection{Studies on ground-state scenario}

Semidefinite relaxations have long been used to investigate ground-state properties of many-body systems~\cite{wang2024certifying, barthel2012solving}. 
We now examine the performance of SDP-assisted overlapping tomography in estimating ground states of local Hamiltonians.

\begin{figure}[h]
    \centering
    \includegraphics[width=1\columnwidth]{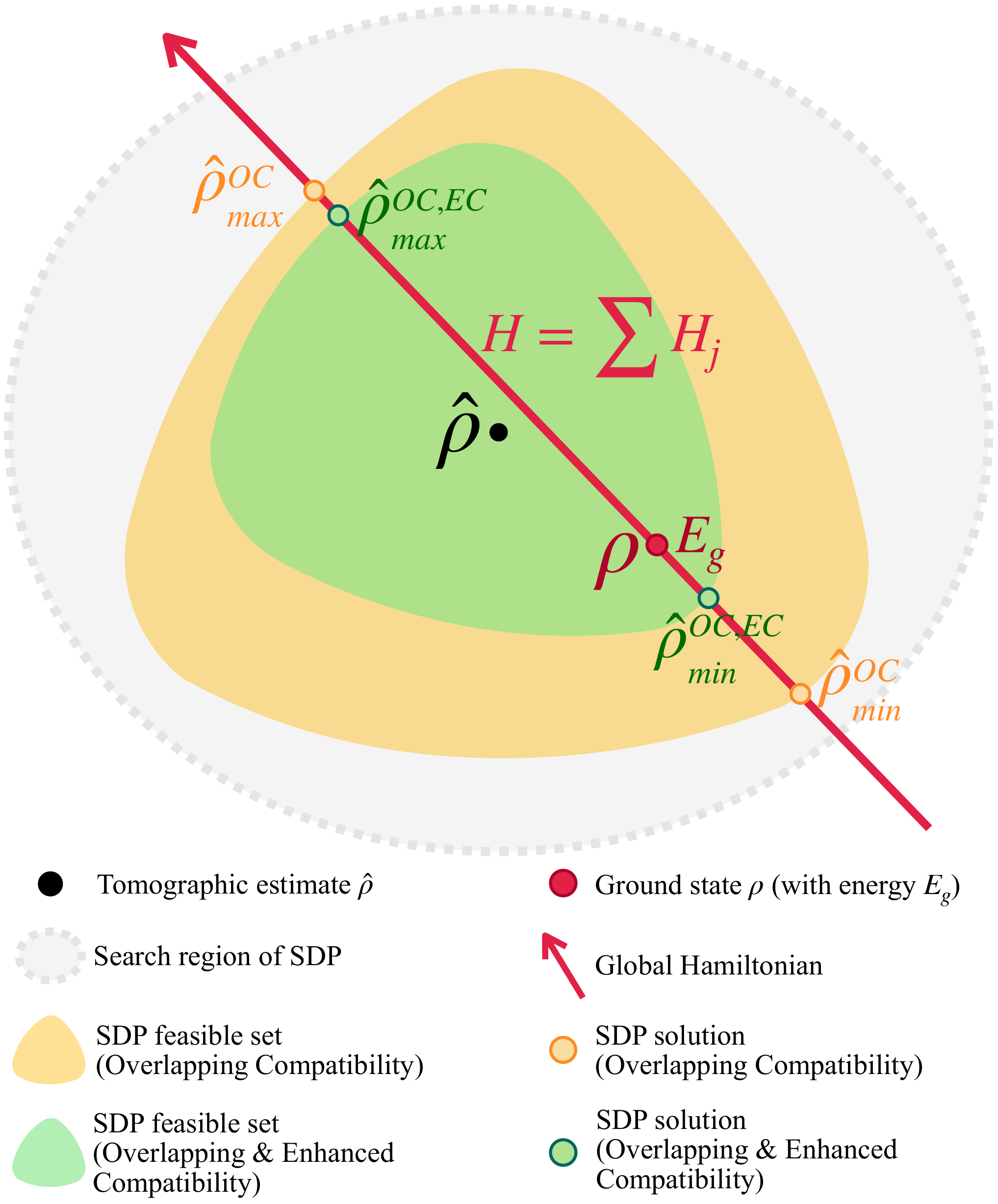}
    \caption{\textbf{Feasible set for the SDP problem.}
    An $n$-qubit system with RDM $\rho$ and energy $E_g$ (shown in red) undergoes tomography, resulting in an approximate RDM $\hat{\rho}$ with energy $\hat{E}$ (black). 
    The shaded gray region represents the (high-dimensional) search space. 
    The yellow spectrahedron depicts the feasible set with only overlapping-compatibility (OC) constraints, whereas the green spectrahedron also includes enhanced-compatibility (EC).}
    \label{fig:epsilon}
\end{figure}

Fig.~\ref{fig:epsilon} illustrates two feasible sets under the same search region but with or without EC constraints. These feasible sets are defined by all RDMs satisfying both the overlapping- (\(\mathscr{R}\)) and enhanced-compatibility (\(\mathscr{G}\)) constraints or only the overlapping (\(\mathscr{R}\)) constraints, respectively, while the search region refers to the intersection of positive semidefinite cones restricted by the confidence intervals in Eq.~\eqref{eq:C_interval}. Introducing EC constraints narrows the feasible set and yields tighter lower and upper bounds on ground-state energy estimate $\hat{E}_g$:
\begin{align}\Tr\left[H\hat{\rho}_{_{\text{min}}}^{_{\text{OC,EC}}}\right] \geq  \Tr\left[H\hat{\rho}_{_{\text{min}}}^{_{\text{OC}}}\right],\\
\Tr\left[H\hat{\rho}_{_{\text{max}}}^{_{\text{OC,EC}}}\right] \leq  \Tr\left[H\hat{\rho}_{_{\text{max}}}^{_{\text{OC}}}\right].
\end{align}

The variable \(\epsilon^j_\mathbf{i} = \alpha\,\mathrm{Var}(\hat{C}^j_{\mathbf{i}})\) in Eq.~\eqref{eq:C_interval} defines the search region explored by the SDP, thereby controlling the confidence interval for the ground-state energy estimate. Different \(\alpha\)-values yield different search regions, visualized in the right panel of Fig.~\ref{fig:SDP_curve}. The spectrahedra illustrates the feasible set after imposing positive semidefiniteness, OC, and EC constraints. The true state \(\rho\) (red) and its estimated counterpart \(\hat{\rho}\) (black) are included for comparison. 
Crucially, when the search region is large enough within the feasible set to capture the true RDMs $\rho$, we obtain upper and lower bounds on the ground energy estimate  $\hat{E}_g$:
\begin{align} \label{eq:inequality}
\mathrm{Tr}\bigl[H\,\hat{\rho}_{\mathrm{min}}^{\mathrm{SDP}}(\alpha_0)\bigr]
\,\le\,
\hat{E}_g
\,\le\,
\mathrm{Tr}\bigl[H\,\hat{\rho}_{\mathrm{max}}^{\mathrm{SDP}}(\alpha_1)\bigr],
\end{align}
where \(\alpha_0>\alpha_1\).

\begin{algorithm}[h]
   \caption{SDP-assisted tomography on ground states\label{algorithm1}}
   \KwIn{$T$ identical ground-state preparations, 
   Hamiltonian $H = \sum_{j=1}^m H_j$,
   hyperparameters $b>\Delta >0$, where $\Delta$ is the tolerance.} 
   \KwOut{Local RDM estimates $\{\hat{\rho}_j^{\text{SDP}}\}_{j=1}^m$ that minimize the energy, and the corresponding minimum energy $\hat{E}_g$.}
   Perform QST to obtain estimated local RDMs $\{\hat{\rho}_j\}_{j=1}^m$ with the precision given by $T$ samples\;
   \Repeat{$\mathcal{F} \neq \emptyset$}{
   Solve the SDP problem described by Eq.\eqref{eq:SDP_problem} with decision variables $\{\Tilde{\rho}_j\}_{j=1}^m$ and coefficient $\alpha = b$\;
   Denote the feasible set as $\mathcal{F}$\;
   $b \gets 2b$ \;
   }
   Initialize $a = 0$\;
   \While{$|a - b| \geq \Delta$}{
       $\alpha_0 \gets a + \frac{|a - b|}{2}$\;
       Solve the SDP problem described by Eq.\eqref{eq:SDP_problem} with decision variables $\{\Tilde{\rho}_j\}_{j=1}^m$ and coefficient $\alpha = \alpha_0$\;
       Denote the feasible set as $\mathcal{F}$\;
       \eIf{$\mathcal{F} \neq \emptyset$}{
           Denote the solution as $\mathcal{S} = \arg\min_{ \{ \Tilde{\rho}_j \} } \sum_j \Tr\left[\Tilde{\rho}_j H_j\right]$\;
           $b \gets \alpha_0$\;
       }{
           $a \gets \alpha_0$\;
       }
   }
   $\{\hat{\rho}_j^{\text{SDP}}\}_{j=1}^m \gets \mathcal{S}$\;
   $\hat{E}_g \gets \sum_{j=1}^{m}\Tr[ \hat{\rho}_j^{\text{SDP}} H_j ]$\;
\end{algorithm}

We employ a bisection method (Alg.~\ref{algorithm1}) to identify the smallest \(\alpha_0\) (within a tolerance \(\Delta_0\)) for which the feasible set is non-empty, and analogously for \(\alpha_1\) using a maximization variant. 
This allows us to establish the bounds of Eq.~\eqref{eq:inequality} on \(\hat{E}_g\).

\subsection{Numerical simulations \label{sec:sdp_result}}

In this section, we present numerical results for estimating the ground-state energy of the 1D-chain \(XY\) model~\cite{lieb1961two}. 
We also compare the estimation accuracy achieved by standard tomography to that of the proposed SDP-assisted tomography.

\subsubsection{Problem description}

The Hamiltonian of the 1D-chain \(XY\) model
is given by
\begin{align}
    H_{C} = \sum_{j=1}^{n-1} H_j = J \sum_{j=1}^{n-1} \left(X_jX_{j+1} + Y_jY_{j+1} \right). \label{eq:h_XY}
\end{align}
where
\(H_j\) is the local two-qubit Hamiltonian term capturing the interaction between qubits \(j\) and \(j+1\). 
Here, \(n\) denotes the total number of qubits, and \(J\) is the interaction strength. The subscript "C" stands for "chain".

Although the accuracy of ground-state energy estimation through tomography is fundamentally limited by the number of ground-state samples, introducing additional constraints can significantly tighten the resulting confidence interval. 
In the context of the \(XY\) model $H_{C}$, the overlapping- (OC) and enhanced-compatibility (EC) constraints in Eq.~\eqref{eq:SDP_problem} take the form
\begin{align}
    \mathscr{R}_{C}\coloneqq &\cup_{j \in [n-1]}\mathcal{R}_{C}\left( \Tilde{\rho}_j, \Tilde{\rho}_{j+1} \right), \\    \mathscr{G}_{C}\coloneqq&\cup_{j \in [n-2]
    }\mathcal{G}_{C}\left( \Tilde{\rho}_j, \Tilde{\rho}_{j+1}\right),
\end{align}
respectively, with
\begin{align}
\mathcal{R}_{C}\left( \Tilde{\rho}_j, \Tilde{\rho}_{j+1} \right) &\coloneqq \{
\Tr_{\{j\}}\left[ \Tilde{\rho}_j \right]-\Tr_{\{j+2\}}\left[ \Tilde{\rho}_{j+1} \right] = 0\},\label{eq:R} \\
\mathcal{G}_{C}\left( \Tilde{\rho}_j, \Tilde{\rho}_{j+1}\right) &\coloneqq
\{
\Tilde{\rho} \succeq 0,  \Tr_{\{j\}}\left[{\Tilde{\rho}}\right] = \Tilde{\rho}_{j+1}, 
\Tr_{\{j+2\}}
\left[{\Tilde{\rho}}\right] = \Tilde{\rho}_{j}
\}.\label{eq:G}
\end{align}
Here, \(\Tilde{\rho}_j\) is the SDP decision variable corresponding to the two-qubit subsystem of qubits \((j,j+1)\), and \(\Tilde{\rho}\) is a three-qubit SDP variable spanning qubits \(j, j+1,\) and \(j+2\). The subscript in the partial trace indicates which qubit is being traced out.

Note that the energy function can be computed from the two-qubit RDMs in a natural way, but also equivalently from the three-qubit RDMs that stem from the ECs, either by taking their partial traces (which are compatible with the two-qubit RDMs, by construction) or by extending the Hamiltonian terms to a larger Hilbert spaces (by appropriately tensoring them with identity operators).

\begin{figure*}[t]
\centering
\includegraphics[width=2.05\columnwidth]{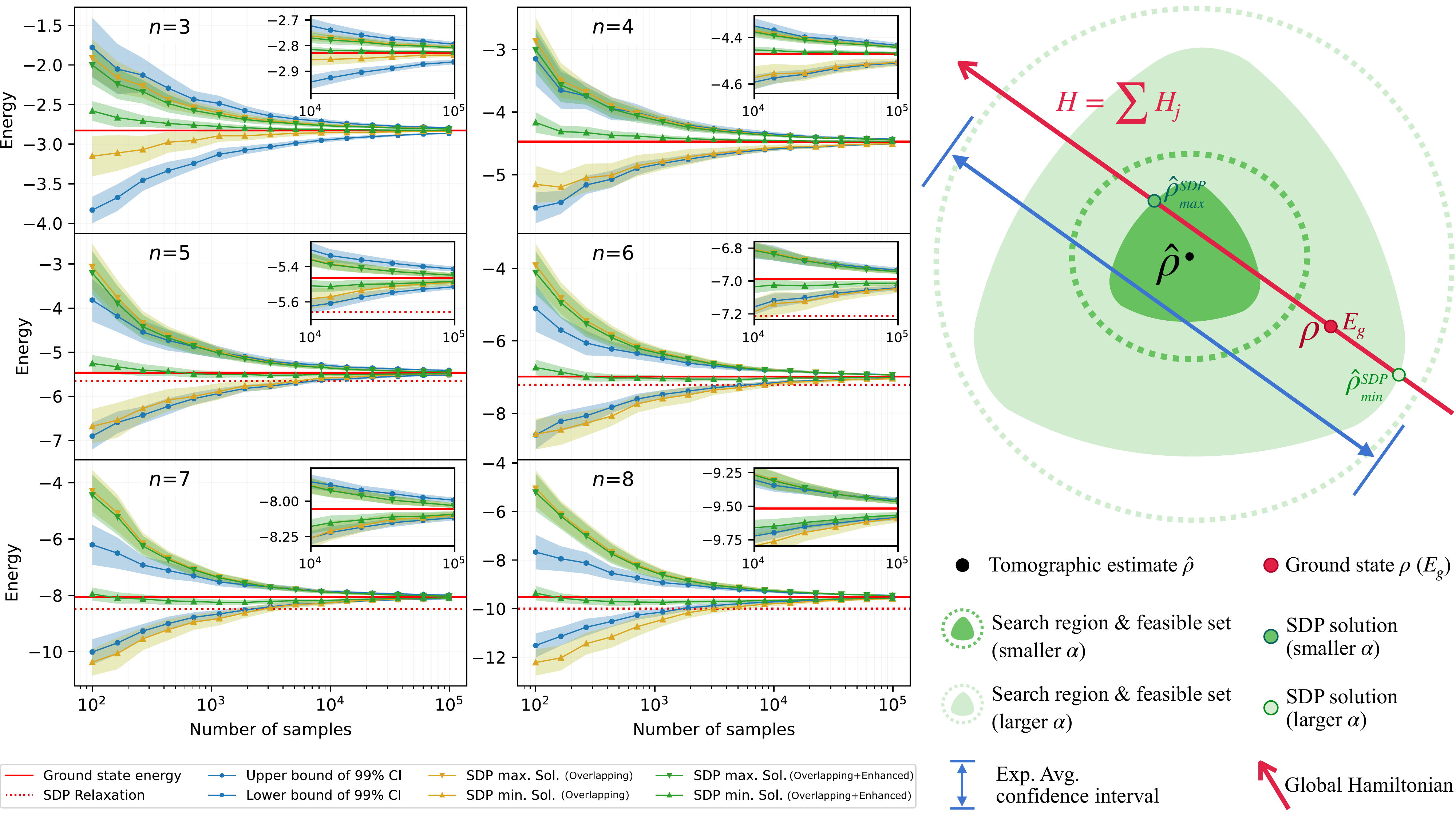}
	\caption{
\textbf{Tightening expectation value confidence interval via SDP-assisted quantum tomography.}
\textit{(Right)} 
An illustrative depiction of the SDP feasibility region. 
The true RDM \(\rho\) with energy \(E_g\) of Hamiltonian $H$ (red) and its tomographic estimate \(\hat{\rho}\) (black) lie within a search region of adjustable size (dashed circle) determined by the coefficient $\alpha$.
The feasible set under overlapping-compatibility (OC) and enhanced-compatibility (EC) constraints appears as a shaded spectrahedron.
The blue interval indicates the confidence interval for the average of the energy obtained by standard quantum state tomography from the simulated data.
\textit{(Left)}
Ground-state energy estimates for the \(XY\) model \((J=1)\) with \(n=3,4,5,6,7,8\) qubits against the number of state samples. 
The red solid lines indicate the exact ground-state energy; the red dashed lines are lower bounds from the relaxation method. 
The blue curves show the \(99\%\) confidence interval upper and lower bounds for standard quantum tomography.
The yellow (green) curves give the results of SDP minimization with  tolerance parameters \(\Delta_0 = 0.1\), and maximization with  tolerance parameters \(\Delta_0 = 0.001\), under OC (OC+EC) constraints, respectively. 
In each plot, the two green curves can be interpreted as the upper and lower energy bounds using SDP-enhanced quantum tomography.}

\label{fig:SDP_curve}
\end{figure*}

\subsubsection{Explanation on the results}

For the SDP-assisted tomography, we are firstly given a number of identical copies of the \(XY\) model ground state, as in Eq.~\eqref{eq:h_XY}. 
We perform a series of measurements in randomly chosen Pauli basis \(\sigma_{\mathbf{i}}\), with \(\mathbf{i}\) uniformly sampled from \(\{1,2,3\}^n\). 
From these measurement outcomes, we reconstruct the local RDMs via standard tomography and estimate the ground-state energy through Eq.~\eqref{eq:exp_local_rdm}. 
We then apply our SDP-based post-processing to refine these ground-state energy estimates.

The plots on the left side of Fig.~\ref{fig:SDP_curve} show the resulting confidence intervals for the ground-state energy as a function of the number of samples, for \(n=3,4,5,6,7,8\). 
In each subplot, the red solid line indicates the exact ground-state energy from Eq.~\eqref{eq:h_XY} obtained by exact diagonalization, while the blue region marks the \(99\%\) confidence interval from standard tomography. 
The green curves represent the solutions of Alg.~\ref{algorithm1} applied to SDP minimization and maximization (with OC+EC constraints, Eqs.~\eqref{eq:R} and~\eqref{eq:G}) and two different tolerances \(\Delta_0=0.1\) and \(\Delta_1=0.001\). 
This choice is motivated by the fact that there is a strong asymmetry between upper and lower bounds when tomographing the ground state.
In contrast, the yellow curves illustrate the SDP results with only OC constraints (Eq.~\eqref{eq:R}).

Comparing the yellow and green curves marked with triangles shows that adding EC constraints substantially tightens the lower bound obtained by SDP minimization. 
When the sample size is small, the tomographic RDM \(\hat{\rho}\) can significantly diverge from the true RDM \(\rho\), creating bias in the SDP energy bound in the presence of large statistical noise. 
 Together with the fact that the probability of failure in estimating the true ground state is higher in extremely-low-shot regimes, this fact makes it much more likely for the green curves marked with triangles to exceed the exact ground-state energy.
However, as the number of samples grows, the SDP-assisted approach narrows the confidence interval more effectively than standard tomography. 
This improvement is especially clear in the zoomed-in subplots, where the green curve, representing SDP-assisted (OC+EC) bounds, spans a narrower range for \(\langle H\rangle\) than the blue curve from standard QST. 
Notably, SDP-assisted tomography can reduce the required number of samples by a factor of $10^{1}$ to $10^{2}$ compared to standard QST, while achieving the same level of precision in lower-bounding the energy.

The red dashed lines represent lower bounds obtained via an existing relaxation method~\cite{barthel2012solving}. 
Such approaches optimize the ground-state energy under only a subset of the constraints that we use, thus guaranteeing a strict, albeit potentially conservative, bound.
While such relaxations have successfully established lower bounds for the ground-state energies of local Hamiltonians in many-body settings \cite{mazziotti2006variational, requena2023certificates, mazziotti2004realization, navascues2007bounding, barthel2012solving, rai2024hierarchy, kull2024lower}, the accuracy of these bounds tends to degrade for larger problem sizes.
This shortfall arises from omitting higher-level EC constraints to keep the computational effort manageable.

All numerical simulations were performed using QISKIT~\cite{javadi2024quantum} for quantum simulations and CVXPY~\cite{agrawal2018rewriting, diamond2016cvxpy} to solve SDP formulations using the SCS optimizer.

\section{Application 
\label{sec:app}}

In this section, we demonstrate how to integrate our SDP-assisted tomography into a variational procedure for preparing and characterizing low-energy states of local Hamiltonians. 
Specifically, we embed SDP-assisted tomography within the \emph{algorithmic cooling} (AC) method~\cite{boykin2002algorithmic, polla2021quantum, grimsley2019adaptive} to heuristically minimize the energy of a target quantum Hamiltonian. 
Section~\ref{sec:ac_intro} introduces AC, and Section~\ref{sec:embed_sdp} explains how SDP-assisted tomography is incorporated into the AC workflow. Numerical results for the 1D-chain \(XY\) model are presented in Section~\ref{sec:ac_results}, comparing the performance of AC both with and without semidefinite programming.

\subsection{Algorithmic cooling\label{sec:ac_intro}}
\begin{figure}[htpb!]
    \centering
    \includegraphics[width=0.98\columnwidth]{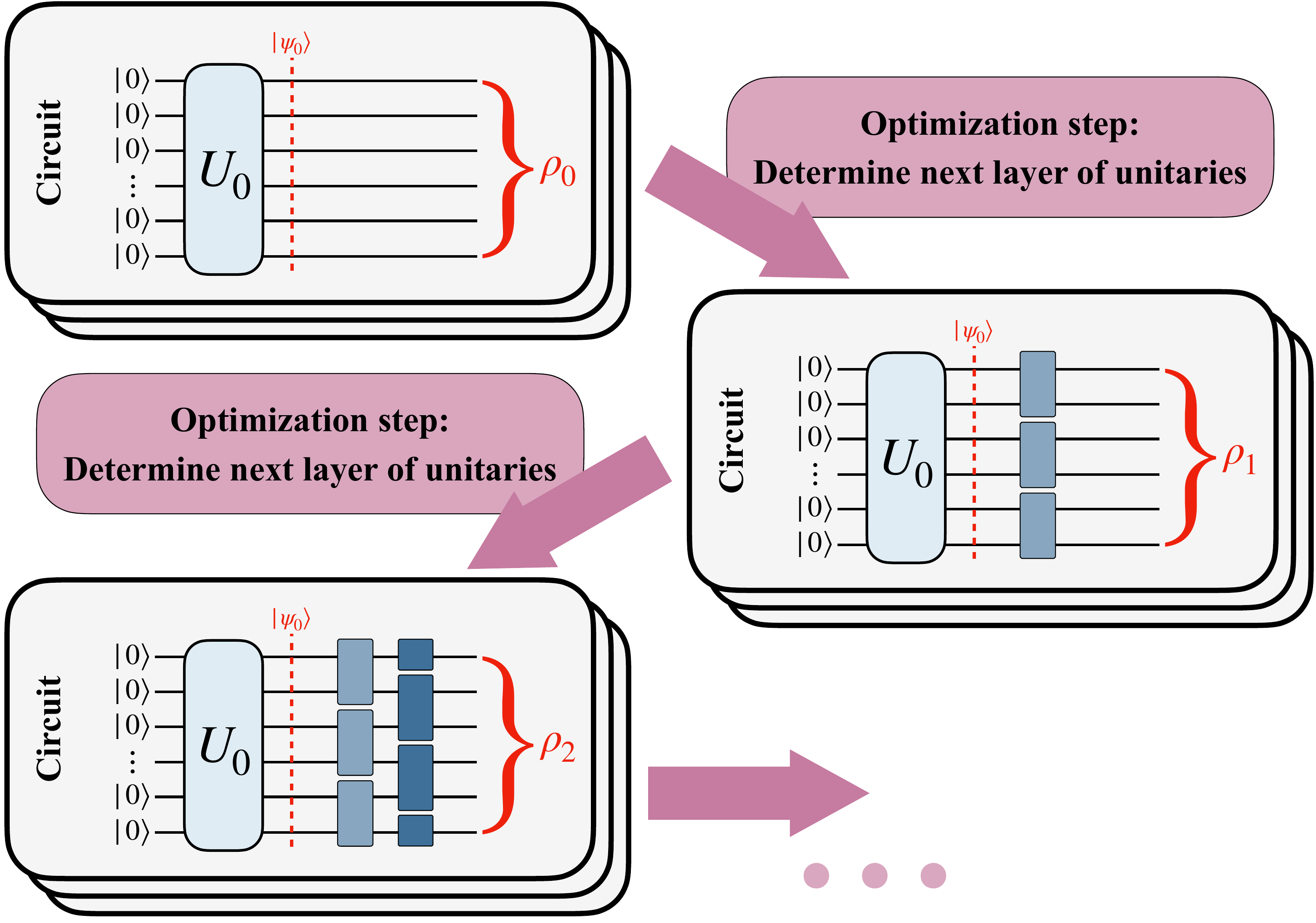}
    \caption{\textbf{Conceptual diagram of algorithmic cooling (AC) \label{fig:AC}}
    Algorithmic cooling iteratively constructs a quantum circuit that prepares a low-energy state of a local Hamiltonian. Initially, the circuit is the identity, and the trial state is \(\ket{\psi_0} = U_0 \ket{0}^{\otimes n}\). Each iteration generates an ensemble of identical states for tomography, yielding local RDMs that guide the choice of a new layer of gates (see Section  \ref{sec:ac_optimization}), which is appended to form the next circuit.}
    \label{fig:ac_illustration}
\end{figure}

Algorithmic cooling (AC) aims to prepare a low-energy state \(\ket{\psi}\) of an \(n\)-qubit quantum system governed by a local Hamiltonian $H$, using near-term quantum devices. 
The AC method is adapted to the practical capabilities of the experimental setup, which can differ considerably across platforms. 
Thus, we do not assume access to arbitrary unitaries; instead, we denote by \(\mathbf{h}\) the set of Hermitian operators \(h\) for which \(e^{-iht}\) can be implemented natively on the device. 
The unitary \(e^{-iht}\) is implemented by turning on an interaction given by $h$ for an amount of time $t$.
We also assume, without loss of generality, that each \(h \in \mathbf{h}\) does not commute with \(H\) (Hamiltonian of \(XY\) model). Fig.~\ref{fig:AC} outlines the AC workflow.

\subsubsection{Optimization step \label{sec:ac_optimization}}

We now describe the optimization routine in AC, which incrementally builds a shallow circuit to prepare a low-energy state for local Hamiltonians. 
We begin by introducing the cooling principle.

Consider a Hamiltonian \(H\) of the form in Eq.~\eqref{eq:H_local} and an operator \(h \in \mathbf{h}\) such that \([H, h] \neq 0\) and, for simplicity, \(h^2 = \mathbb{I}\). 
The last assumption can be relaxed~\cite{markovich2024parameter}. 
Given an initial quantum state \(\ket{\psi_0}\),
\begin{align}
\left|\psi_0^h(t)\right\rangle =& e^{-i t h}\left|\psi_0\right\rangle\\
    =& \cos(t)\mathbb{I}-i\sin(t)h \left|\psi_0\right\rangle, \notag
\end{align}
we define the energy function:
\begin{align} E_h(t)=&\left\langle\psi_0^h(t)\right| H\left|\psi_0^h(t)\right\rangle\\
    =& \cos^2(t)\langle H\rangle + i \cos(t) \sin(t)\langle[h, H]\rangle + \sin^2(t)\langle h H h\rangle. \notag
\end{align} 
By using trigonometric identities, one finds
\begin{align}
    E_h(t) 
    &= \langle H\rangle
    \;+\;
    A\,\sin^2(t) 
    \;+\;
    \tfrac{B}{2}\,\sin(2t),
\end{align}
where \(\langle \cdot \rangle\) indicates the expectation value in \(\ket{\psi_0}\), \(A \coloneqq \langle h H h - H\rangle\), and \(B \coloneqq i \langle [h, H]\rangle\).
The minimum of \(E_h(t)\) occurs at
\begin{align}
    t^*=\frac{1}{2} \arctan \frac{-B}{A}, \label{eq:t*}
\end{align}
yielding
\begin{align}\label{eq:energy_decrease}
E_h\left(t^*\right)=\langle H\rangle+\frac{1}{2}\left(A-\sqrt{A^2+B^2}\right)\ .
\end{align}

Note that there is a slight ambiguity in the definition of $t^*$ given in Eq.~\eqref{eq:t*}: the minimum is achieved at any $t^*$ satisfying $\cos \left(2 t^*\right)=A / \sqrt{A^2+B^2}$ and $\sin \left(2 t^*\right)=-B / \sqrt{A^2+B^2}$. 
Choosing the opposite sign leads to a maximum at $t^* + \pi/2$,
\begin{align}
E_h\left(t^*+\pi/2\right)=\langle H\rangle+\left(A+\sqrt{A^2+B^2}\right)/2.
\end{align}

Therefore, if \(A\) and \(B\) can be accurately estimated, applying 
 the interaction given by $h$ (or $-h$) for a period of time $|t^*|$ will decrease the energy of $\ket{\psi_0}$ by an amount of $\frac{1}{2}\left(\sqrt{A^2+B^2}-A\right)$.
Since \(H\) is local, the value of \(A\) and \(B\) can be extracted from measurements on only a constant number of qubits:
\begin{align}
    A=&\sum_{j: \operatorname{supp}\left(H_j \cap h\right) \neq \emptyset}\left\langle h H_j h-H_j\right\rangle, \label{eq:A}\\
    B=&\sum_{j: \operatorname{supp}\left(H_j \cap h\right) \neq \emptyset} i\left\langle\left[h, H_j\right]\right\rangle. \label{eq:B}
\end{align}
Here, $\operatorname{supp}(H_j \cap h) \coloneqq \operatorname{supp}(H_j) \cap \operatorname{supp}(h)$, and \(\mathrm{supp}(\cdot)\subseteq [n]\) denotes the qubits on which an operator acts non-trivially.

Because each update step is chosen to locally decrease energy, the energy function forms a monotonically decreasing sequence (assuming we have access to exact RDMs $\{\rho_j\}_{j=1}^m$) that is bounded by below, guaranteeing convergence (albeit not necessarily to the true ground state). 
The algorithmic cooling approach is thus a heuristic strategy that can fail for several reasons: \(\mathbf{h}\) may be too small to reach the global ground state from the initial state; choosing unitaries in certain orders could lead to local minima; or noisy RDM estimates might produce suboptimal energy updates.

Nonetheless, AC provides a flexible, greedy technique for variationally reducing energy. 
Integrating SDP-assisted tomography can enhance the accuracy of local RDM estimates, thereby improving the reliability and efficiency of each iteration.

\subsubsection{Algorithmic cooling circuit compilation}

This subsection explains how we construct the layout of the variational circuit by integrating the optimization step from Section~\ref{sec:ac_optimization} into the algorithmic cooling (AC) method. Note that there are many ways to construct the heuristic, each with a different performance depending on the problem. Here, we outline a specific one for illustrative purposes.

We begin with an empty circuit, $\mathcal{U}_C = \mathbb{I}$, and an initial trial state $\ket{\psi_0}$. 
At each step, we consider all available operators $h \in \mathbf{h}$, selecting the one that yields the greatest energy reduction.
We then append the corresponding unitary $U_k = e^{-i h_k t_k^*}$ to the circuit, ensuring a strictly decreasing energy at each iteration. 
Although this approach guarantees monotonic improvement, it may not fully exploit device capabilities. 

To make better use of circuit depth, we group gates into layers $V_1, \dots, V_L$. Each layer
$V_l = U_l^1 \cdots U_l^k \cdots U_l^{K_l}$ consists of $K_l$ parallel unitaries, with disjoint supports,
\begin{align}
\operatorname{supp}\left(U_l^k\right) \cap \operatorname{supp}\left(U_l^{k'}\right) = \emptyset, \quad
    \forall \ 1 \leq k, k' \leq K_l.
\end{align}
The overall circuit is then $\mathcal{U}_C = V_L \cdots V_1$, allowing for a more efficient arrangement while still achieving a steady decrease in energy. 

To optimize a layer $V_l$, one could iteratively:
\begin{enumerate}
    \item Select an operator $h$, compute $t^*$, and update $V_l \leftarrow V_l\,e^{-iht^*}$.
    \item Repeat for the next operator $h'$, finding $t'^*$, and so on.
\end{enumerate}
However, each new operator choice requires a full state preparation and measurement cycle. Since $H$ is local, it can become more resource-efficient to collect a suitable set of RDMs for the current layout of $V_l$ and then perform all unitary updates for that layer classically, rather than measuring after every single addition of $e^{-iht^*}$.

In Alg.~\ref{alg:ac}, we present a more refined approach that avoids predefining the circuit layout. 
Instead, at each iteration, the algorithm greedily selects the operator $h^*$ and corresponding $t^*$ that produce the largest energy decrease. 
The selection is restricted to operators whose supports are disjoint from any non-trivial gates already in the layer. 
This ensures the circuit grows incrementally while maintaining a clear and consistent reduction in energy at each step.

\begin{algorithm}[h]
   \caption{Algorithmic cooling \label{alg:ac}}
   \KwIn{Initial state $\ket{\psi_0}$, Hamiltonian $H=\sum_{j=1}^m H_j$, operator set $\mathbf{h}_{0}$, maximum number of iterations $L$.}
   \KwOut{Quantum circuit $\mathcal{U}_C$.}
   Initialize the system to $\ket{\psi_0}$ and set $\mathcal{U}_C = \mathbb{I}$\;
   $l \gets 0$\;
   \Repeat{$l = L$}{
      Prepare an ensemble of identical states $\mathcal{U}_C\ket{\psi_0}$ and perform QST to obtain local RDMs $\{\hat{\rho}_j\}$\;
      $\mathbf{h} \gets \mathbf{h}_{0}$, $V_l \gets \mathbb{I}$\;
      \While{$\mathbf{h} \neq \emptyset$}{
         Estimate the parameters $A$ (Eq.~\eqref{eq:A}) and $B$ (Eq.~\eqref{eq:B}) for every $h \in \mathbf{h}$\ with local RDMs $\{\hat{\rho}_j\}$, respectively\;
         Select the $h^*$ and $t^*$ that maximize the energy decrease $\frac{1}{2}\left(\sqrt{A^2+B^2}-A\right)$, c.f. Eq.~\eqref{eq:energy_decrease}\;
         $V_l \gets e^{-ih^* t^*} V_l$\;
         $\hat{\rho}_j \gets e^{-ih^* t^*} \hat{\rho}_j e^{i(h^*)^{\dagger} t^*},\ \forall \hat{\rho}_j\in \{\hat{\rho}_j\}$\;
         $\mathbf{h} \gets \mathbf{h} \setminus \{ h \in \mathbf{h} \mid \operatorname{supp}(h) \cap \operatorname{supp}(h^*) \neq \emptyset \}$\;
      }
      $\mathcal{U}_C \gets  V_l\mathcal{U}_C$\;
      $l \gets l + 1$\;
   }
\end{algorithm}

\subsection{Embedding SDP-assisted tomography \label{sec:embed_sdp}}

In practice, the set of possible operators \(\mathbf{h}\) can be extremely large. 
Rather than testing each \(h \in \mathbf{h}\) individually, one can instead perform tomography on the reduced density matrices (RDMs) of relevant qubit subsets. 
Although tomography itself is expensive, if both the Hamiltonian \(H\) and the unitaries to be optimized are local, then the quantities 
\(\langle h H_j h - H_j\rangle\) and \(\langle [h, H_j]\rangle\)
needed to identify the optimal unitary \(e^{-iht^*}\) depend only on the RDMs of
\(\mathrm{supp}(H_j) \cup \mathrm{supp}(h)\), which is constant for every $j$. 
These expectation values from Eqs.~\eqref{eq:A} and \eqref{eq:B} can be computed via
\begin{align}
    \langle h H_j h - H_j\rangle 
    &= 
    \mathrm{Tr}\bigl[\rho_r \bigl(h H_j h - H_j\bigr)\bigr],\\
    i\langle [h, H_j]\rangle 
    &= 
    i\,\mathrm{Tr}\bigl[\rho_r\,[h, H_j]\bigr],
\end{align}
where \(\rho_r\) is a collection of relevant RDMs of the current global state \(\ket{\psi}\):
\begin{align}
    \rho_r = \Tr_{[n] \setminus \left(\operatorname{supp}\left(H_j\right) \cup \operatorname{supp}\left(h\right)\right)} (\ketbra{\psi}{\psi}).
\end{align}

To obtain the relevant RDMs, one first performs state tomography.
However, due to shot noise, the raw tomographic local RDMs may become non-physical. 
To address this, one can formulate an SDP (Eq.~\eqref{eq:SDP_problem}) to recover locally consistent RDMs that minimize the energy while preserving physicality.
These SDP-refined RDMs are then provided to the algorithmic cooling (AC) method (Alg.~\ref{alg:ac}), guiding the optimization parameters for each cooling step.

\subsection{Numerical simulations \label{sec:ac_results}}

We now present numerical simulations of the AC procedure for approximating the ground state of the 1D-chain \(XY\) model, specified by Eq.~\eqref{eq:h_XY}. 
We consider two initial states: (a) the uniform superposition \(\ket{+}^{\otimes n}\) and (b) a Hartree--Fock (mean-field) product state, which corresponds to the product state of minimal energy, which can be obtained efficiently in one dimension \cite{schuch2010matrix}, thereby providing an in principle better initialization. 
We restrict the set of available operators $\mathbf{h}$ to geometrically local Pauli operators.

\begin{figure*}[htpb!]
\centering
\includegraphics[width=2.05\columnwidth]{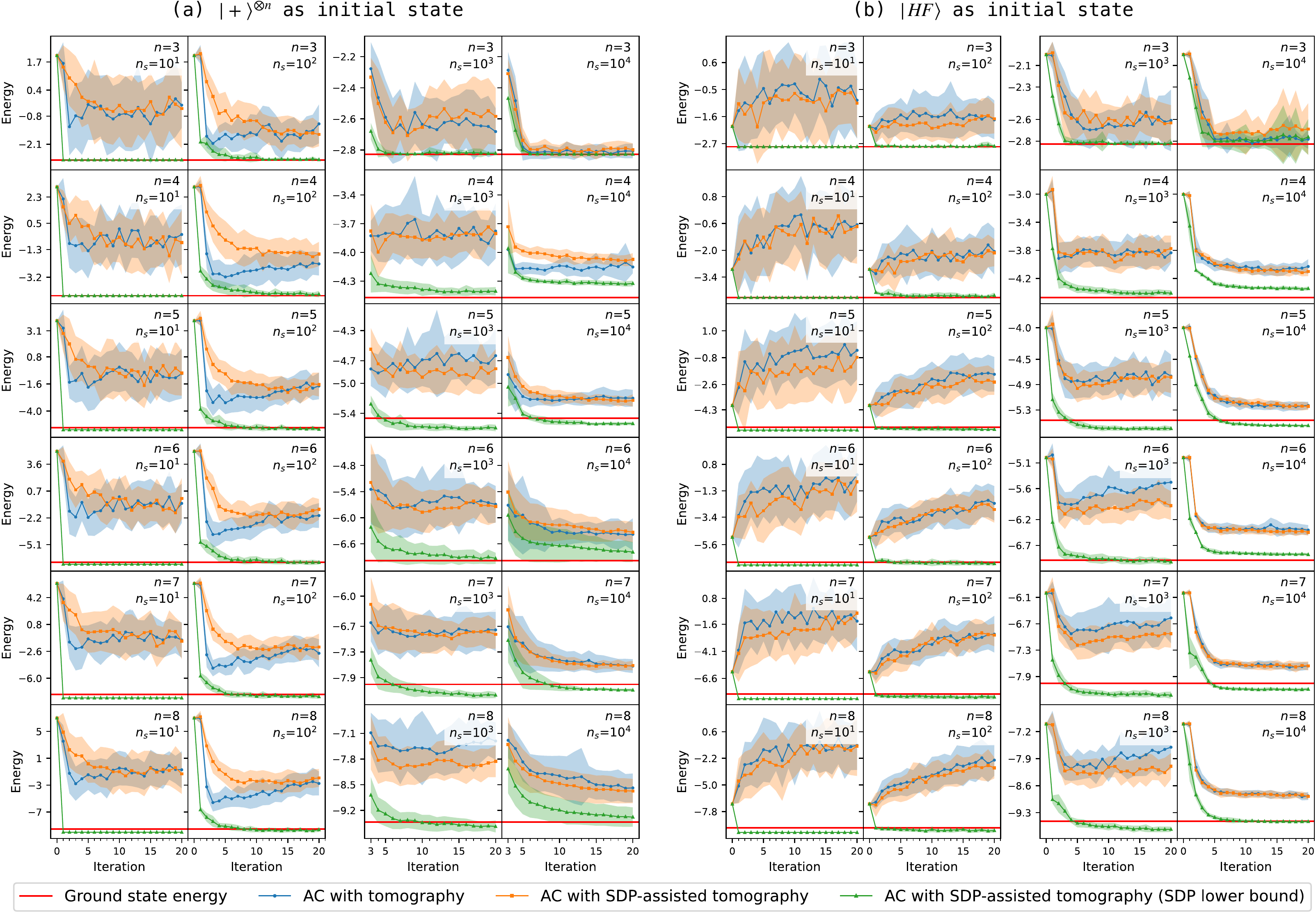}
	\caption{\label{fig:ac_result}
\textbf{Algorithmic cooling for ground-state preparation of a 1D-chain \(XY\) model (frustrated Hamiltonian).}
We employ Alg.~\ref{alg:ac} that iteratively prepares the ground state, starting from either \(\ket{+}^{\otimes n}\) (left panel) or a Hartree-Fock state (right panel). 
Each iteration uses $n_s$ samples per iteration to estimate the necessary parameters for cooling either using naive tomography or SDP-assisted tomography.
The subplots show energy expectation values as a function of AC iterations for $n=3,4,5,6,7,8$ qubits and $n_s=10^1,10^2,10^3,10^4$ samples per iteration.
The red line indicates the exact ground-state energy; 
the blue curve represents AC using naive tomography, 
the orange curve uses SDP-refined local RDMs for the energy expectation, 
and the green curve displays the minimal energy compatible with the measured data under SDP constraints (i.e., the SDP lower bound).
The shaded regions show one-sided standard deviation over 25 runs.
}
\label{fig:ac_frustrated}
\end{figure*}

Fig.~\ref{fig:ac_result} illustrates the results for different pairs \((n,n_s)\), where \(n\) (number of qubits) ranges from 3 to 8 and \(n_s\) (number of samples per iteration) takes values \(10^1, 10^2, 10^3, 10^4\). 
Within each subplot, the red line marks the exact ground-state energy.
Two variants of AC method are simulated: (i) the blue curve corresponds to AC with naive tomography, using tomographically reconstructed local RDMs to compute energy, and 
(ii) the orange curve depicts AC with SDP-refined RDMs, still using their reconstructed expectation value for energy.
The green curve shows the SDP lower-bound on the energy at each iteration, representing the minimal energy compatible with the measured data under the chosen constraints.
Each method is repeated 25 times, and the shaded areas around the curves indicate one-sided standard deviation.

Fig. \ref{fig:ac_result} also showcases three noteworthy regimes: 
For a low number of measurements $(\sim 10^2)$, the monotonicity property, especially of the AC with tomography, is broken. 
This is due to the shot noise being too dominant and confusing the heuristic of the AC. 
For the $\ket{+}^{\otimes n}$ as an initial state, and for $n_{s}=10^2$, we observe a sudden dip in energy for the AC with tomography, followed by a steady increase in energy, as iterations progress. 
This is due to the AC adding gates that are not sufficiently close to optimal, because of the limitations in precision.
In contrast, the behavior of AC with SDP, is qualitatively closer to monotonicity, especially in the first few iterations, although ultimately converging to a similar value as the AC with tomography.
For the $\ket{HF}$ as initial state, both AC with tomography and AC with SDP have the same phenomena, but the energy scales are much lower as we begin from the lowest-energy product state.

For a higher number of measurements $(\sim 10^3)$, the AC with SDP performs better compared to the AC with tomography. 
This showcases a sweet-spot regime where our approach proves more advantageous, as this behavior is consistent across different system sizes and both initializations considered. 
However, for an even larger number of measurements ($\sim 10^4$), the performance of the two methods becomes similar. 
This is likely because, with more measurement data, the estimates provided by the tomography are increasingly physical, leaving less room for the SDP to correct non-physicality. 
In the absence of shot noise, the performance of both methods is identical by construction.

It is worth noting that the number of samples used in our numerical simulations is significantly lower than in typical physical experiments, which highlights the efficiency of the approach, though it introduces some limitations. 
It should be noted that there is no guaranteed advantage of the orange curve (AC with SDP-optimized local RDMs) over the blue curve (AC with local RDMs obtained via tomography) in any single simulation.
This is because AC is a heuristic method, and its performance depends on various factors, such as the initial state, measurement settings, and other parameters. 
However, in some settings, the blue and orange curves have non-overlapping shaded areas.
Additionally, by construction, the green curve (AC with SDP-optimized local RDMs and minimized energy) yields a lower bound on the orange curve.

\section{Conclusion and outlook \label{sec:outlook}}

In this work, we introduced a semidefinite programming (SDP)-assisted technique for reconstructing overlapping reduced density matrices (RDMs) from experimental measurement data, effectively addressing the challenges posed by shot noise in near-term quantum computing. 
By enforcing overlapping-compatibility (OC) and enhanced-compatibility (EC) constraints on local RDMs through a polynomial-sized SDP problem, our method alleviates the impact of limited measurements and ensures the consistency of the reconstructed local RDMs. 
As a result, our approach not only enhances the accuracy of local observable estimates but also provides tighter confidence intervals compared to standard tomographic procedures.

To illustrate its effectiveness, we applied SDP-assisted tomography within a variational quantum algorithm, algorithmic cooling (AC), aimed at heuristically preparing low-energy states of local Hamiltonians. 
Numerical simulations on 1D-chain \(XY\) model demonstrated the advantages of our approach in terms of both accuracy and resource efficiency. 
These findings indicate that SDP-assisted tomography can serve as a valuable asset for boosting the performance of variational quantum algorithms and other quantum information processing tasks in the near term.

\vspace{10pt}
Looking ahead, a promising direction for future research is the incorporation of additional constraints, such as entropy constraints, into the SDP formulation to further refine the reconstruction process \cite{fawzi2023entropy}. 
It would also be interesting to extend the SDP-assisted tomography framework to more complex many-body systems and investigate its behavior on frustration-free Hamiltonians. 
Applying it to specific quantum chemistry calculations
might be promising, as done in a similar way in Refs.~\cite{anselmetti2024classical,avdic2024fewer}. 
In addition, exploiting the framework to optimize other local observables, such as correlation functions, is another potential area of exploration. 
 Another open question is the effect of systematic errors on the application of our method, and how strategies such as those presented in  \cite{PhysRevA.86.062325} can help to reduce them.

Finally, examining how SDP-based methods can be combined with noise mitigation strategies on quantum hardware could offer valuable insights for practical enhancements in near-term quantum computing.

\section{Acknowledgments \label{sec:acknowledgments}}
The authors thank I. Cirac, B. Terhal, D. Elkouss, N. Liu, M. Gärttner, J.F. Chen, Y. Herasymenko, S. Polla, A. Shankar, K. S. Rai, M. Stroeks,  N. Euler and M. Reckermann for useful discussions during the project.
Z.J.W., D.D., and J.T. acknowledge the support received from the Dutch National Growth Fund (NGF), as part of the Quantum Delta NL programme.
J.T. acknowledges the support received from the European Union’s Horizon Europe research and innovation programme through the ERC StG FINE-TEA-SQUAD (Grant No. 101040729).
This publication is part of the ‘Quantum Inspire — the Dutch Quantum Computer in the Cloud’ project (NWA.1292.19.194) of the NWA research program ‘Research on Routes by Consortia (ORC)’, which is funded by the Netherlands Organization for Scientific Research (NWO).
The views and opinions expressed here are solely those of the authors and do not necessarily reflect those of the funding institutions. Neither of the funding institution can be held responsible for them. 
\section{Code Availability}
The code supporting this work is available upon reasonable request.

\bibliography{bib}

\providecommand{\noopsort}[1]{}\providecommand{\singleletter}[1]{#1}%
\begin{thebibliography}{51}%
\makeatletter
\providecommand \@ifxundefined [1]{%
 \@ifx{#1\undefined}
}%
\providecommand \@ifnum [1]{%
 \ifnum #1\expandafter \@firstoftwo
 \else \expandafter \@secondoftwo
 \fi
}%
\providecommand \@ifx [1]{%
 \ifx #1\expandafter \@firstoftwo
 \else \expandafter \@secondoftwo
 \fi
}%
\providecommand \natexlab [1]{#1}%
\providecommand \enquote  [1]{``#1''}%
\providecommand \bibnamefont  [1]{#1}%
\providecommand \bibfnamefont [1]{#1}%
\providecommand \citenamefont [1]{#1}%
\providecommand \href@noop [0]{\@secondoftwo}%
\providecommand \href [0]{\begingroup \@sanitize@url \@href}%
\providecommand \@href[1]{\@@startlink{#1}\@@href}%
\providecommand \@@href[1]{\endgroup#1\@@endlink}%
\providecommand \@sanitize@url [0]{\catcode `\\12\catcode `\$12\catcode `\&12\catcode `\#12\catcode `\^12\catcode `\_12\catcode `\%12\relax}%
\providecommand \@@startlink[1]{}%
\providecommand \@@endlink[0]{}%
\providecommand \url  [0]{\begingroup\@sanitize@url \@url }%
\providecommand \@url [1]{\endgroup\@href {#1}{\urlprefix }}%
\providecommand \urlprefix  [0]{URL }%
\providecommand \Eprint [0]{\href }%
\providecommand \doibase [0]{https://doi.org/}%
\providecommand \selectlanguage [0]{\@gobble}%
\providecommand \bibinfo  [0]{\@secondoftwo}%
\providecommand \bibfield  [0]{\@secondoftwo}%
\providecommand \translation [1]{[#1]}%
\providecommand \BibitemOpen [0]{}%
\providecommand \bibitemStop [0]{}%
\providecommand \bibitemNoStop [0]{.\EOS\space}%
\providecommand \EOS [0]{\spacefactor3000\relax}%
\providecommand \BibitemShut  [1]{\csname bibitem#1\endcsname}%
\let\auto@bib@innerbib\@empty
\bibitem [{\citenamefont {Nielsen}\ and\ \citenamefont {Chuang}(2010)}]{nielsen2010quantum}%
  \BibitemOpen
  \bibfield  {author} {\bibinfo {author} {\bibfnamefont {M.~A.}\ \bibnamefont {Nielsen}}\ and\ \bibinfo {author} {\bibfnamefont {I.~L.}\ \bibnamefont {Chuang}},\ }\href@noop {} {\emph {\bibinfo {title} {Quantum computation and quantum information}}}\ (\bibinfo  {publisher} {Cambridge university press},\ \bibinfo {year} {2010})\BibitemShut {NoStop}%
\bibitem [{\citenamefont {Lvovsky}\ and\ \citenamefont {Raymer}(2009)}]{lvovsky2009continuous}%
  \BibitemOpen
  \bibfield  {author} {\bibinfo {author} {\bibfnamefont {A.~I.}\ \bibnamefont {Lvovsky}}\ and\ \bibinfo {author} {\bibfnamefont {M.~G.}\ \bibnamefont {Raymer}},\ }\bibfield  {title} {\bibinfo {title} {Continuous-variable optical quantum-state tomography},\ }\href@noop {} {\bibfield  {journal} {\bibinfo  {journal} {Reviews of modern physics}\ }\textbf {\bibinfo {volume} {81}},\ \bibinfo {pages} {299} (\bibinfo {year} {2009})}\BibitemShut {NoStop}%
\bibitem [{\citenamefont {Cramer}\ \emph {et~al.}(2010)\citenamefont {Cramer}, \citenamefont {Plenio}, \citenamefont {Flammia}, \citenamefont {Somma}, \citenamefont {Gross}, \citenamefont {Bartlett}, \citenamefont {Landon-Cardinal}, \citenamefont {Poulin},\ and\ \citenamefont {Liu}}]{cramer2010efficient}%
  \BibitemOpen
  \bibfield  {author} {\bibinfo {author} {\bibfnamefont {M.}~\bibnamefont {Cramer}}, \bibinfo {author} {\bibfnamefont {M.~B.}\ \bibnamefont {Plenio}}, \bibinfo {author} {\bibfnamefont {S.~T.}\ \bibnamefont {Flammia}}, \bibinfo {author} {\bibfnamefont {R.}~\bibnamefont {Somma}}, \bibinfo {author} {\bibfnamefont {D.}~\bibnamefont {Gross}}, \bibinfo {author} {\bibfnamefont {S.~D.}\ \bibnamefont {Bartlett}}, \bibinfo {author} {\bibfnamefont {O.}~\bibnamefont {Landon-Cardinal}}, \bibinfo {author} {\bibfnamefont {D.}~\bibnamefont {Poulin}},\ and\ \bibinfo {author} {\bibfnamefont {Y.-K.}\ \bibnamefont {Liu}},\ }\bibfield  {title} {\bibinfo {title} {Efficient quantum state tomography},\ }\href@noop {} {\bibfield  {journal} {\bibinfo  {journal} {Nature communications}\ }\textbf {\bibinfo {volume} {1}},\ \bibinfo {pages} {149} (\bibinfo {year} {2010})}\BibitemShut {NoStop}%
\bibitem [{\citenamefont {Kliesch}\ and\ \citenamefont {Roth}(2021)}]{PRXQuantum.2.010201}%
  \BibitemOpen
  \bibfield  {author} {\bibinfo {author} {\bibfnamefont {M.}~\bibnamefont {Kliesch}}\ and\ \bibinfo {author} {\bibfnamefont {I.}~\bibnamefont {Roth}},\ }\bibfield  {title} {\bibinfo {title} {Theory of quantum system certification},\ }\href {https://doi.org/10.1103/PRXQuantum.2.010201} {\bibfield  {journal} {\bibinfo  {journal} {PRX Quantum}\ }\textbf {\bibinfo {volume} {2}},\ \bibinfo {pages} {010201} (\bibinfo {year} {2021})}\BibitemShut {NoStop}%
\bibitem [{\citenamefont {Cerezo}\ \emph {et~al.}(2021)\citenamefont {Cerezo}, \citenamefont {Arrasmith}, \citenamefont {Babbush}, \citenamefont {Benjamin}, \citenamefont {Endo}, \citenamefont {Fujii}, \citenamefont {McClean}, \citenamefont {Mitarai}, \citenamefont {Yuan}, \citenamefont {Cincio} \emph {et~al.}}]{cerezo2021variational}%
  \BibitemOpen
  \bibfield  {author} {\bibinfo {author} {\bibfnamefont {M.}~\bibnamefont {Cerezo}}, \bibinfo {author} {\bibfnamefont {A.}~\bibnamefont {Arrasmith}}, \bibinfo {author} {\bibfnamefont {R.}~\bibnamefont {Babbush}}, \bibinfo {author} {\bibfnamefont {S.~C.}\ \bibnamefont {Benjamin}}, \bibinfo {author} {\bibfnamefont {S.}~\bibnamefont {Endo}}, \bibinfo {author} {\bibfnamefont {K.}~\bibnamefont {Fujii}}, \bibinfo {author} {\bibfnamefont {J.~R.}\ \bibnamefont {McClean}}, \bibinfo {author} {\bibfnamefont {K.}~\bibnamefont {Mitarai}}, \bibinfo {author} {\bibfnamefont {X.}~\bibnamefont {Yuan}}, \bibinfo {author} {\bibfnamefont {L.}~\bibnamefont {Cincio}}, \emph {et~al.},\ }\bibfield  {title} {\bibinfo {title} {Variational quantum algorithms},\ }\href@noop {} {\bibfield  {journal} {\bibinfo  {journal} {Nature Reviews Physics}\ }\textbf {\bibinfo {volume} {3}},\ \bibinfo {pages} {625} (\bibinfo {year} {2021})}\BibitemShut {NoStop}%
\bibitem [{\citenamefont {Huang}\ \emph {et~al.}(2020)\citenamefont {Huang}, \citenamefont {Kueng},\ and\ \citenamefont {Preskill}}]{Huang_2020}%
  \BibitemOpen
  \bibfield  {author} {\bibinfo {author} {\bibfnamefont {H.-Y.}\ \bibnamefont {Huang}}, \bibinfo {author} {\bibfnamefont {R.}~\bibnamefont {Kueng}},\ and\ \bibinfo {author} {\bibfnamefont {J.}~\bibnamefont {Preskill}},\ }\bibfield  {title} {\bibinfo {title} {Predicting many properties of a quantum system from very few measurements},\ }\href {https://doi.org/10.1038/s41567-020-0932-7} {\bibfield  {journal} {\bibinfo  {journal} {Nature Physics}\ }\textbf {\bibinfo {volume} {16}},\ \bibinfo {pages} {1050–1057} (\bibinfo {year} {2020})}\BibitemShut {NoStop}%
\bibitem [{\citenamefont {Liu}\ \emph {et~al.}(2021)\citenamefont {Liu}, \citenamefont {Li},\ and\ \citenamefont {Yang}}]{liu2021efficient}%
  \BibitemOpen
  \bibfield  {author} {\bibinfo {author} {\bibfnamefont {J.}~\bibnamefont {Liu}}, \bibinfo {author} {\bibfnamefont {Z.}~\bibnamefont {Li}},\ and\ \bibinfo {author} {\bibfnamefont {J.}~\bibnamefont {Yang}},\ }\bibfield  {title} {\bibinfo {title} {An efficient adaptive variational quantum solver of the schr{\"o}dinger equation based on reduced density matrices},\ }\href@noop {} {\bibfield  {journal} {\bibinfo  {journal} {The Journal of chemical physics}\ }\textbf {\bibinfo {volume} {154}} (\bibinfo {year} {2021})}\BibitemShut {NoStop}%
\bibitem [{\citenamefont {Ara{\'u}jo}\ \emph {et~al.}(2022)\citenamefont {Ara{\'u}jo}, \citenamefont {Taddei}, \citenamefont {Cavalcanti},\ and\ \citenamefont {Ac{\'\i}n}}]{araujo2022local}%
  \BibitemOpen
  \bibfield  {author} {\bibinfo {author} {\bibfnamefont {B.~G.}\ \bibnamefont {Ara{\'u}jo}}, \bibinfo {author} {\bibfnamefont {M.~M.}\ \bibnamefont {Taddei}}, \bibinfo {author} {\bibfnamefont {D.}~\bibnamefont {Cavalcanti}},\ and\ \bibinfo {author} {\bibfnamefont {A.}~\bibnamefont {Ac{\'\i}n}},\ }\bibfield  {title} {\bibinfo {title} {Local quantum overlapping tomography},\ }\href@noop {} {\bibfield  {journal} {\bibinfo  {journal} {Physical Review A}\ }\textbf {\bibinfo {volume} {106}},\ \bibinfo {pages} {062441} (\bibinfo {year} {2022})}\BibitemShut {NoStop}%
\bibitem [{\citenamefont {Davidson}(2012)}]{davidson2012reduced}%
  \BibitemOpen
  \bibfield  {author} {\bibinfo {author} {\bibfnamefont {E.}~\bibnamefont {Davidson}},\ }\href@noop {} {\emph {\bibinfo {title} {Reduced density matrices in quantum chemistry}}},\ Vol.~\bibinfo {volume} {6}\ (\bibinfo  {publisher} {Elsevier},\ \bibinfo {year} {2012})\BibitemShut {NoStop}%
\bibitem [{\citenamefont {Linden}\ \emph {et~al.}(2002)\citenamefont {Linden}, \citenamefont {Popescu},\ and\ \citenamefont {Wootters}}]{linden2002almost}%
  \BibitemOpen
  \bibfield  {author} {\bibinfo {author} {\bibfnamefont {N.}~\bibnamefont {Linden}}, \bibinfo {author} {\bibfnamefont {S.}~\bibnamefont {Popescu}},\ and\ \bibinfo {author} {\bibfnamefont {W.}~\bibnamefont {Wootters}},\ }\bibfield  {title} {\bibinfo {title} {Almost every pure state of three qubits is completely determined<? format?> by its two-particle reduced density matrices},\ }\href@noop {} {\bibfield  {journal} {\bibinfo  {journal} {Physical review letters}\ }\textbf {\bibinfo {volume} {89}},\ \bibinfo {pages} {207901} (\bibinfo {year} {2002})}\BibitemShut {NoStop}%
\bibitem [{\citenamefont {Linden}\ and\ \citenamefont {Wootters}(2002)}]{linden2002parts}%
  \BibitemOpen
  \bibfield  {author} {\bibinfo {author} {\bibfnamefont {N.}~\bibnamefont {Linden}}\ and\ \bibinfo {author} {\bibfnamefont {W.}~\bibnamefont {Wootters}},\ }\bibfield  {title} {\bibinfo {title} {The parts determine the whole in a generic pure quantum state},\ }\href@noop {} {\bibfield  {journal} {\bibinfo  {journal} {Physical review letters}\ }\textbf {\bibinfo {volume} {89}},\ \bibinfo {pages} {277906} (\bibinfo {year} {2002})}\BibitemShut {NoStop}%
\bibitem [{\citenamefont {Cotler}\ and\ \citenamefont {Wilczek}(2020)}]{cotler2020quantum}%
  \BibitemOpen
  \bibfield  {author} {\bibinfo {author} {\bibfnamefont {J.}~\bibnamefont {Cotler}}\ and\ \bibinfo {author} {\bibfnamefont {F.}~\bibnamefont {Wilczek}},\ }\bibfield  {title} {\bibinfo {title} {Quantum overlapping tomography},\ }\href@noop {} {\bibfield  {journal} {\bibinfo  {journal} {Physical review letters}\ }\textbf {\bibinfo {volume} {124}},\ \bibinfo {pages} {100401} (\bibinfo {year} {2020})}\BibitemShut {NoStop}%
\bibitem [{\citenamefont {Klyachko}(2006)}]{klyachko2006quantum}%
  \BibitemOpen
  \bibfield  {author} {\bibinfo {author} {\bibfnamefont {A.~A.}\ \bibnamefont {Klyachko}},\ }\bibfield  {title} {\bibinfo {title} {Quantum marginal problem and n-representability},\ }\href@noop {} {\bibfield  {journal} {\bibinfo  {journal} {Journal of Physics: Conference Series}\ }\textbf {\bibinfo {volume} {36}},\ \bibinfo {pages} {72} (\bibinfo {year} {2006})}\BibitemShut {NoStop}%
\bibitem [{\citenamefont {Broadbent}\ and\ \citenamefont {Grilo}(2022)}]{broadbent2022qma}%
  \BibitemOpen
  \bibfield  {author} {\bibinfo {author} {\bibfnamefont {A.}~\bibnamefont {Broadbent}}\ and\ \bibinfo {author} {\bibfnamefont {A.~B.}\ \bibnamefont {Grilo}},\ }\bibfield  {title} {\bibinfo {title} {Qma-hardness of consistency of local density matrices with applications to quantum zero-knowledge},\ }\href@noop {} {\bibfield  {journal} {\bibinfo  {journal} {SIAM Journal on Computing}\ }\textbf {\bibinfo {volume} {51}},\ \bibinfo {pages} {1400} (\bibinfo {year} {2022})}\BibitemShut {NoStop}%
\bibitem [{\citenamefont {Kamminga}\ and\ \citenamefont {Rudolph}(2024)}]{kamminga2024complexitypurestateconsistencylocal}%
  \BibitemOpen
  \bibfield  {author} {\bibinfo {author} {\bibfnamefont {J.}~\bibnamefont {Kamminga}}\ and\ \bibinfo {author} {\bibfnamefont {D.}~\bibnamefont {Rudolph}},\ }\href {https://arxiv.org/abs/2411.03096} {\bibinfo {title} {On the complexity of pure-state consistency of local density matrices}} (\bibinfo {year} {2024}),\ \Eprint {https://arxiv.org/abs/2411.03096} {arXiv:2411.03096 [quant-ph]} \BibitemShut {NoStop}%
\bibitem [{\citenamefont {McCaskey}\ \emph {et~al.}(2019)\citenamefont {McCaskey}, \citenamefont {Parks}, \citenamefont {Jakowski}, \citenamefont {Moore}, \citenamefont {Morris}, \citenamefont {Humble},\ and\ \citenamefont {Pooser}}]{mccaskey2019quantum}%
  \BibitemOpen
  \bibfield  {author} {\bibinfo {author} {\bibfnamefont {A.~J.}\ \bibnamefont {McCaskey}}, \bibinfo {author} {\bibfnamefont {Z.~P.}\ \bibnamefont {Parks}}, \bibinfo {author} {\bibfnamefont {J.}~\bibnamefont {Jakowski}}, \bibinfo {author} {\bibfnamefont {S.~V.}\ \bibnamefont {Moore}}, \bibinfo {author} {\bibfnamefont {T.~D.}\ \bibnamefont {Morris}}, \bibinfo {author} {\bibfnamefont {T.~S.}\ \bibnamefont {Humble}},\ and\ \bibinfo {author} {\bibfnamefont {R.~C.}\ \bibnamefont {Pooser}},\ }\bibfield  {title} {\bibinfo {title} {Quantum chemistry as a benchmark for near-term quantum computers},\ }\href@noop {} {\bibfield  {journal} {\bibinfo  {journal} {npj Quantum Information}\ }\textbf {\bibinfo {volume} {5}},\ \bibinfo {pages} {99} (\bibinfo {year} {2019})}\BibitemShut {NoStop}%
\bibitem [{\citenamefont {Hansenne}\ \emph {et~al.}(2024)\citenamefont {Hansenne}, \citenamefont {Qu}, \citenamefont {Weinbrenner}, \citenamefont {de~Gois}, \citenamefont {Wang}, \citenamefont {Ming}, \citenamefont {Yang}, \citenamefont {Horodecki}, \citenamefont {Gao},\ and\ \citenamefont {G{\"u}hne}}]{hansenne2024optimal}%
  \BibitemOpen
  \bibfield  {author} {\bibinfo {author} {\bibfnamefont {K.}~\bibnamefont {Hansenne}}, \bibinfo {author} {\bibfnamefont {R.}~\bibnamefont {Qu}}, \bibinfo {author} {\bibfnamefont {L.~T.}\ \bibnamefont {Weinbrenner}}, \bibinfo {author} {\bibfnamefont {C.}~\bibnamefont {de~Gois}}, \bibinfo {author} {\bibfnamefont {H.}~\bibnamefont {Wang}}, \bibinfo {author} {\bibfnamefont {Y.}~\bibnamefont {Ming}}, \bibinfo {author} {\bibfnamefont {Z.}~\bibnamefont {Yang}}, \bibinfo {author} {\bibfnamefont {P.}~\bibnamefont {Horodecki}}, \bibinfo {author} {\bibfnamefont {W.}~\bibnamefont {Gao}},\ and\ \bibinfo {author} {\bibfnamefont {O.}~\bibnamefont {G{\"u}hne}},\ }\bibfield  {title} {\bibinfo {title} {Optimal overlapping tomography},\ }\href@noop {} {\bibfield  {journal} {\bibinfo  {journal} {arXiv preprint arXiv:2408.05730}\ } (\bibinfo {year} {2024})}\BibitemShut {NoStop}%
\bibitem [{\citenamefont {Skrzypczyk}\ and\ \citenamefont {Cavalcanti}(2023)}]{skrzypczyk2023semidefinite}%
  \BibitemOpen
  \bibfield  {author} {\bibinfo {author} {\bibfnamefont {P.}~\bibnamefont {Skrzypczyk}}\ and\ \bibinfo {author} {\bibfnamefont {D.}~\bibnamefont {Cavalcanti}},\ }\href@noop {} {\emph {\bibinfo {title} {Semidefinite Programming in Quantum Information Science}}}\ (\bibinfo  {publisher} {IOP Publishing},\ \bibinfo {year} {2023})\BibitemShut {NoStop}%
\bibitem [{\citenamefont {Aloy}\ \emph {et~al.}(2021)\citenamefont {Aloy}, \citenamefont {Fadel},\ and\ \citenamefont {Tura}}]{aloy2021quantum}%
  \BibitemOpen
  \bibfield  {author} {\bibinfo {author} {\bibfnamefont {A.}~\bibnamefont {Aloy}}, \bibinfo {author} {\bibfnamefont {M.}~\bibnamefont {Fadel}},\ and\ \bibinfo {author} {\bibfnamefont {J.}~\bibnamefont {Tura}},\ }\bibfield  {title} {\bibinfo {title} {The quantum marginal problem for symmetric states: applications to variational optimization, nonlocality and self-testing},\ }\href@noop {} {\bibfield  {journal} {\bibinfo  {journal} {New Journal of Physics}\ }\textbf {\bibinfo {volume} {23}},\ \bibinfo {pages} {033026} (\bibinfo {year} {2021})}\BibitemShut {NoStop}%
\bibitem [{\citenamefont {Schwemmer}\ \emph {et~al.}(2015)\citenamefont {Schwemmer}, \citenamefont {Knips}, \citenamefont {Richart}, \citenamefont {Weinfurter}, \citenamefont {Moroder}, \citenamefont {Kleinmann},\ and\ \citenamefont {G{\"u}hne}}]{schwemmer2015systematic}%
  \BibitemOpen
  \bibfield  {author} {\bibinfo {author} {\bibfnamefont {C.}~\bibnamefont {Schwemmer}}, \bibinfo {author} {\bibfnamefont {L.}~\bibnamefont {Knips}}, \bibinfo {author} {\bibfnamefont {D.}~\bibnamefont {Richart}}, \bibinfo {author} {\bibfnamefont {H.}~\bibnamefont {Weinfurter}}, \bibinfo {author} {\bibfnamefont {T.}~\bibnamefont {Moroder}}, \bibinfo {author} {\bibfnamefont {M.}~\bibnamefont {Kleinmann}},\ and\ \bibinfo {author} {\bibfnamefont {O.}~\bibnamefont {G{\"u}hne}},\ }\bibfield  {title} {\bibinfo {title} {Systematic errors in current quantum state tomography tools},\ }\href@noop {} {\bibfield  {journal} {\bibinfo  {journal} {Physical review letters}\ }\textbf {\bibinfo {volume} {114}},\ \bibinfo {pages} {080403} (\bibinfo {year} {2015})}\BibitemShut {NoStop}%
\bibitem [{\citenamefont {Boykin}\ \emph {et~al.}(2002)\citenamefont {Boykin}, \citenamefont {Mor}, \citenamefont {Roychowdhury}, \citenamefont {Vatan},\ and\ \citenamefont {Vrijen}}]{boykin2002algorithmic}%
  \BibitemOpen
  \bibfield  {author} {\bibinfo {author} {\bibfnamefont {P.~O.}\ \bibnamefont {Boykin}}, \bibinfo {author} {\bibfnamefont {T.}~\bibnamefont {Mor}}, \bibinfo {author} {\bibfnamefont {V.}~\bibnamefont {Roychowdhury}}, \bibinfo {author} {\bibfnamefont {F.}~\bibnamefont {Vatan}},\ and\ \bibinfo {author} {\bibfnamefont {R.}~\bibnamefont {Vrijen}},\ }\bibfield  {title} {\bibinfo {title} {Algorithmic cooling and scalable nmr quantum computers},\ }\href@noop {} {\bibfield  {journal} {\bibinfo  {journal} {Proceedings of the National Academy of Sciences}\ }\textbf {\bibinfo {volume} {99}},\ \bibinfo {pages} {3388} (\bibinfo {year} {2002})}\BibitemShut {NoStop}%
\bibitem [{\citenamefont {Polla}\ \emph {et~al.}(2021)\citenamefont {Polla}, \citenamefont {Herasymenko},\ and\ \citenamefont {O'Brien}}]{polla2021quantum}%
  \BibitemOpen
  \bibfield  {author} {\bibinfo {author} {\bibfnamefont {S.}~\bibnamefont {Polla}}, \bibinfo {author} {\bibfnamefont {Y.}~\bibnamefont {Herasymenko}},\ and\ \bibinfo {author} {\bibfnamefont {T.~E.}\ \bibnamefont {O'Brien}},\ }\bibfield  {title} {\bibinfo {title} {Quantum digital cooling},\ }\href@noop {} {\bibfield  {journal} {\bibinfo  {journal} {Physical Review A}\ }\textbf {\bibinfo {volume} {104}},\ \bibinfo {pages} {012414} (\bibinfo {year} {2021})}\BibitemShut {NoStop}%
\bibitem [{\citenamefont {Grimsley}\ \emph {et~al.}(2019)\citenamefont {Grimsley}, \citenamefont {Economou}, \citenamefont {Barnes},\ and\ \citenamefont {Mayhall}}]{grimsley2019adaptive}%
  \BibitemOpen
  \bibfield  {author} {\bibinfo {author} {\bibfnamefont {H.~R.}\ \bibnamefont {Grimsley}}, \bibinfo {author} {\bibfnamefont {S.~E.}\ \bibnamefont {Economou}}, \bibinfo {author} {\bibfnamefont {E.}~\bibnamefont {Barnes}},\ and\ \bibinfo {author} {\bibfnamefont {N.~J.}\ \bibnamefont {Mayhall}},\ }\bibfield  {title} {\bibinfo {title} {An adaptive variational algorithm for exact molecular simulations on a quantum computer},\ }\href@noop {} {\bibfield  {journal} {\bibinfo  {journal} {Nature communications}\ }\textbf {\bibinfo {volume} {10}},\ \bibinfo {pages} {3007} (\bibinfo {year} {2019})}\BibitemShut {NoStop}%
\bibitem [{\citenamefont {Altepeter}\ \emph {et~al.}(2004)\citenamefont {Altepeter}, \citenamefont {James},\ and\ \citenamefont {Kwiat}}]{altepeter20044}%
  \BibitemOpen
  \bibfield  {author} {\bibinfo {author} {\bibfnamefont {J.~B.}\ \bibnamefont {Altepeter}}, \bibinfo {author} {\bibfnamefont {D.~F.}\ \bibnamefont {James}},\ and\ \bibinfo {author} {\bibfnamefont {P.~G.}\ \bibnamefont {Kwiat}},\ }\bibfield  {title} {\bibinfo {title} {4 qubit quantum state tomography},\ }\href@noop {} {\bibfield  {journal} {\bibinfo  {journal} {Quantum state estimation}\ ,\ \bibinfo {pages} {113}} (\bibinfo {year} {2004})}\BibitemShut {NoStop}%
\bibitem [{\citenamefont {{Google}}(2024)}]{sycamoredatasheet}%
  \BibitemOpen
  \bibfield  {author} {\bibinfo {author} {\bibnamefont {{Google}}},\ }\href {https://quantumai.google/hardware/datasheet/weber.pdf} {\bibinfo {title} {Sycamore data sheet}} (\bibinfo {year} {2024}),\ \bibinfo {note} {accessed: 2024-07-22}\BibitemShut {NoStop}%
\bibitem [{\citenamefont {Torlai}\ \emph {et~al.}(2018)\citenamefont {Torlai}, \citenamefont {Mazzola}, \citenamefont {Carrasquilla}, \citenamefont {Troyer}, \citenamefont {Melko},\ and\ \citenamefont {Carleo}}]{torlai2018neural}%
  \BibitemOpen
  \bibfield  {author} {\bibinfo {author} {\bibfnamefont {G.}~\bibnamefont {Torlai}}, \bibinfo {author} {\bibfnamefont {G.}~\bibnamefont {Mazzola}}, \bibinfo {author} {\bibfnamefont {J.}~\bibnamefont {Carrasquilla}}, \bibinfo {author} {\bibfnamefont {M.}~\bibnamefont {Troyer}}, \bibinfo {author} {\bibfnamefont {R.}~\bibnamefont {Melko}},\ and\ \bibinfo {author} {\bibfnamefont {G.}~\bibnamefont {Carleo}},\ }\bibfield  {title} {\bibinfo {title} {Neural-network quantum state tomography},\ }\href@noop {} {\bibfield  {journal} {\bibinfo  {journal} {Nature physics}\ }\textbf {\bibinfo {volume} {14}},\ \bibinfo {pages} {447} (\bibinfo {year} {2018})}\BibitemShut {NoStop}%
\bibitem [{\citenamefont {Karmarkar}(1984)}]{karmarkar1984new}%
  \BibitemOpen
  \bibfield  {author} {\bibinfo {author} {\bibfnamefont {N.}~\bibnamefont {Karmarkar}},\ }\bibfield  {title} {\bibinfo {title} {A new polynomial-time algorithm for linear programming},\ }in\ \href@noop {} {\emph {\bibinfo {booktitle} {Proceedings of the sixteenth annual ACM symposium on Theory of computing}}}\ (\bibinfo {year} {1984})\ pp.\ \bibinfo {pages} {302--311}\BibitemShut {NoStop}%
\bibitem [{\citenamefont {Jiang}\ \emph {et~al.}(2020)\citenamefont {Jiang}, \citenamefont {Kathuria}, \citenamefont {Lee}, \citenamefont {Padmanabhan},\ and\ \citenamefont {Song}}]{jiang2020faster}%
  \BibitemOpen
  \bibfield  {author} {\bibinfo {author} {\bibfnamefont {H.}~\bibnamefont {Jiang}}, \bibinfo {author} {\bibfnamefont {T.}~\bibnamefont {Kathuria}}, \bibinfo {author} {\bibfnamefont {Y.~T.}\ \bibnamefont {Lee}}, \bibinfo {author} {\bibfnamefont {S.}~\bibnamefont {Padmanabhan}},\ and\ \bibinfo {author} {\bibfnamefont {Z.}~\bibnamefont {Song}},\ }\bibfield  {title} {\bibinfo {title} {A faster interior point method for semidefinite programming},\ }in\ \href@noop {} {\emph {\bibinfo {booktitle} {2020 IEEE 61st annual symposium on foundations of computer science (FOCS)}}}\ (\bibinfo {organization} {IEEE},\ \bibinfo {year} {2020})\ pp.\ \bibinfo {pages} {910--918}\BibitemShut {NoStop}%
\bibitem [{\citenamefont {Requena}\ \emph {et~al.}(2023)\citenamefont {Requena}, \citenamefont {Mu{\~n}oz-Gil}, \citenamefont {Lewenstein}, \citenamefont {Dunjko},\ and\ \citenamefont {Tura}}]{requena2023certificates}%
  \BibitemOpen
  \bibfield  {author} {\bibinfo {author} {\bibfnamefont {B.}~\bibnamefont {Requena}}, \bibinfo {author} {\bibfnamefont {G.}~\bibnamefont {Mu{\~n}oz-Gil}}, \bibinfo {author} {\bibfnamefont {M.}~\bibnamefont {Lewenstein}}, \bibinfo {author} {\bibfnamefont {V.}~\bibnamefont {Dunjko}},\ and\ \bibinfo {author} {\bibfnamefont {J.}~\bibnamefont {Tura}},\ }\bibfield  {title} {\bibinfo {title} {Certificates of quantum many-body properties assisted by machine learning},\ }\href@noop {} {\bibfield  {journal} {\bibinfo  {journal} {Physical Review Research}\ }\textbf {\bibinfo {volume} {5}},\ \bibinfo {pages} {013097} (\bibinfo {year} {2023})}\BibitemShut {NoStop}%
\bibitem [{\citenamefont {Bonet-Monroig}\ \emph {et~al.}(2020)\citenamefont {Bonet-Monroig}, \citenamefont {Babbush},\ and\ \citenamefont {O’Brien}}]{bonet2020nearly}%
  \BibitemOpen
  \bibfield  {author} {\bibinfo {author} {\bibfnamefont {X.}~\bibnamefont {Bonet-Monroig}}, \bibinfo {author} {\bibfnamefont {R.}~\bibnamefont {Babbush}},\ and\ \bibinfo {author} {\bibfnamefont {T.~E.}\ \bibnamefont {O’Brien}},\ }\bibfield  {title} {\bibinfo {title} {Nearly optimal measurement scheduling for partial tomography of quantum states},\ }\href@noop {} {\bibfield  {journal} {\bibinfo  {journal} {Physical Review X}\ }\textbf {\bibinfo {volume} {10}},\ \bibinfo {pages} {031064} (\bibinfo {year} {2020})}\BibitemShut {NoStop}%
\bibitem [{\citenamefont {de~Almeida}\ \emph {et~al.}(2023)\citenamefont {de~Almeida}, \citenamefont {Kleinmann},\ and\ \citenamefont {Sent{\'\i}s}}]{de2023comparison}%
  \BibitemOpen
  \bibfield  {author} {\bibinfo {author} {\bibfnamefont {J.~O.}\ \bibnamefont {de~Almeida}}, \bibinfo {author} {\bibfnamefont {M.}~\bibnamefont {Kleinmann}},\ and\ \bibinfo {author} {\bibfnamefont {G.}~\bibnamefont {Sent{\'\i}s}},\ }\bibfield  {title} {\bibinfo {title} {Comparison of confidence regions for quantum state tomography},\ }\href@noop {} {\bibfield  {journal} {\bibinfo  {journal} {arXiv preprint arXiv:2303.07136}\ } (\bibinfo {year} {2023})}\BibitemShut {NoStop}%
\bibitem [{\citenamefont {Westerheim}\ \emph {et~al.}(2023)\citenamefont {Westerheim}, \citenamefont {Chen}, \citenamefont {Holmes}, \citenamefont {Luo}, \citenamefont {Nuradha}, \citenamefont {Patel}, \citenamefont {Rethinasamy}, \citenamefont {Wang},\ and\ \citenamefont {Wilde}}]{westerheim2023dual}%
  \BibitemOpen
  \bibfield  {author} {\bibinfo {author} {\bibfnamefont {H.}~\bibnamefont {Westerheim}}, \bibinfo {author} {\bibfnamefont {J.}~\bibnamefont {Chen}}, \bibinfo {author} {\bibfnamefont {Z.}~\bibnamefont {Holmes}}, \bibinfo {author} {\bibfnamefont {I.}~\bibnamefont {Luo}}, \bibinfo {author} {\bibfnamefont {T.}~\bibnamefont {Nuradha}}, \bibinfo {author} {\bibfnamefont {D.}~\bibnamefont {Patel}}, \bibinfo {author} {\bibfnamefont {S.}~\bibnamefont {Rethinasamy}}, \bibinfo {author} {\bibfnamefont {K.}~\bibnamefont {Wang}},\ and\ \bibinfo {author} {\bibfnamefont {M.~M.}\ \bibnamefont {Wilde}},\ }\bibfield  {title} {\bibinfo {title} {Dual-vqe: A quantum algorithm to lower bound the ground-state energy},\ }\href@noop {} {\bibfield  {journal} {\bibinfo  {journal} {arXiv preprint arXiv:2312.03083}\ } (\bibinfo {year} {2023})}\BibitemShut {NoStop}%
\bibitem [{\citenamefont {Zambrano}\ \emph {et~al.}(2023)\citenamefont {Zambrano}, \citenamefont {Farina}, \citenamefont {Pagliaro}, \citenamefont {Taddei},\ and\ \citenamefont {Acin}}]{zambrano2023certification}%
  \BibitemOpen
  \bibfield  {author} {\bibinfo {author} {\bibfnamefont {L.}~\bibnamefont {Zambrano}}, \bibinfo {author} {\bibfnamefont {D.}~\bibnamefont {Farina}}, \bibinfo {author} {\bibfnamefont {E.}~\bibnamefont {Pagliaro}}, \bibinfo {author} {\bibfnamefont {M.~M.}\ \bibnamefont {Taddei}},\ and\ \bibinfo {author} {\bibfnamefont {A.}~\bibnamefont {Acin}},\ }\bibfield  {title} {\bibinfo {title} {Certification of quantum state functions under partial information},\ }\href@noop {} {\bibfield  {journal} {\bibinfo  {journal} {arXiv preprint arXiv:2311.06094}\ } (\bibinfo {year} {2023})}\BibitemShut {NoStop}%
\bibitem [{\citenamefont {Wang}\ \emph {et~al.}(2024)\citenamefont {Wang}, \citenamefont {Surace}, \citenamefont {Fr{\'e}rot}, \citenamefont {Legat}, \citenamefont {Renou}, \citenamefont {Magron},\ and\ \citenamefont {Ac{\'\i}n}}]{wang2024certifying}%
  \BibitemOpen
  \bibfield  {author} {\bibinfo {author} {\bibfnamefont {J.}~\bibnamefont {Wang}}, \bibinfo {author} {\bibfnamefont {J.}~\bibnamefont {Surace}}, \bibinfo {author} {\bibfnamefont {I.}~\bibnamefont {Fr{\'e}rot}}, \bibinfo {author} {\bibfnamefont {B.}~\bibnamefont {Legat}}, \bibinfo {author} {\bibfnamefont {M.-O.}\ \bibnamefont {Renou}}, \bibinfo {author} {\bibfnamefont {V.}~\bibnamefont {Magron}},\ and\ \bibinfo {author} {\bibfnamefont {A.}~\bibnamefont {Ac{\'\i}n}},\ }\bibfield  {title} {\bibinfo {title} {Certifying ground-state properties of many-body systems},\ }\href@noop {} {\bibfield  {journal} {\bibinfo  {journal} {Physical Review X}\ }\textbf {\bibinfo {volume} {14}},\ \bibinfo {pages} {031006} (\bibinfo {year} {2024})}\BibitemShut {NoStop}%
\bibitem [{\citenamefont {Kull}\ \emph {et~al.}(2024)\citenamefont {Kull}, \citenamefont {Schuch}, \citenamefont {Dive},\ and\ \citenamefont {Navascu{\'e}s}}]{kull2024lower}%
  \BibitemOpen
  \bibfield  {author} {\bibinfo {author} {\bibfnamefont {I.}~\bibnamefont {Kull}}, \bibinfo {author} {\bibfnamefont {N.}~\bibnamefont {Schuch}}, \bibinfo {author} {\bibfnamefont {B.}~\bibnamefont {Dive}},\ and\ \bibinfo {author} {\bibfnamefont {M.}~\bibnamefont {Navascu{\'e}s}},\ }\bibfield  {title} {\bibinfo {title} {Lower bounds on ground-state energies of local hamiltonians through the renormalization group},\ }\href@noop {} {\bibfield  {journal} {\bibinfo  {journal} {Physical Review X}\ }\textbf {\bibinfo {volume} {14}},\ \bibinfo {pages} {021008} (\bibinfo {year} {2024})}\BibitemShut {NoStop}%
\bibitem [{\citenamefont {Fawzi}\ \emph {et~al.}(2023)\citenamefont {Fawzi}, \citenamefont {Fawzi},\ and\ \citenamefont {Scalet}}]{fawzi2023entropy}%
  \BibitemOpen
  \bibfield  {author} {\bibinfo {author} {\bibfnamefont {H.}~\bibnamefont {Fawzi}}, \bibinfo {author} {\bibfnamefont {O.}~\bibnamefont {Fawzi}},\ and\ \bibinfo {author} {\bibfnamefont {S.~O.}\ \bibnamefont {Scalet}},\ }\href@noop {} {\bibinfo {title} {Entropy constraints for ground energy optimization}} (\bibinfo {year} {2023}),\ \Eprint {https://arxiv.org/abs/2305.06855} {arXiv:2305.06855 [quant-ph]} \BibitemShut {NoStop}%
\bibitem [{\citenamefont {Barthel}\ and\ \citenamefont {H{\"u}bener}(2012)}]{barthel2012solving}%
  \BibitemOpen
  \bibfield  {author} {\bibinfo {author} {\bibfnamefont {T.}~\bibnamefont {Barthel}}\ and\ \bibinfo {author} {\bibfnamefont {R.}~\bibnamefont {H{\"u}bener}},\ }\bibfield  {title} {\bibinfo {title} {Solving condensed-matter ground-state problems by semidefinite relaxations},\ }\href@noop {} {\bibfield  {journal} {\bibinfo  {journal} {Physical review letters}\ }\textbf {\bibinfo {volume} {108}},\ \bibinfo {pages} {200404} (\bibinfo {year} {2012})}\BibitemShut {NoStop}%
\bibitem [{\citenamefont {Liu}\ \emph {et~al.}(2007)\citenamefont {Liu}, \citenamefont {Christandl},\ and\ \citenamefont {Verstraete}}]{liu2007quantum}%
  \BibitemOpen
  \bibfield  {author} {\bibinfo {author} {\bibfnamefont {Y.-K.}\ \bibnamefont {Liu}}, \bibinfo {author} {\bibfnamefont {M.}~\bibnamefont {Christandl}},\ and\ \bibinfo {author} {\bibfnamefont {F.}~\bibnamefont {Verstraete}},\ }\bibfield  {title} {\bibinfo {title} {Quantum computational complexity of the n-representability problem: Qma complete},\ }\href@noop {} {\bibfield  {journal} {\bibinfo  {journal} {Physical review letters}\ }\textbf {\bibinfo {volume} {98}},\ \bibinfo {pages} {110503} (\bibinfo {year} {2007})}\BibitemShut {NoStop}%
\bibitem [{\citenamefont {Lieb}\ \emph {et~al.}(1961)\citenamefont {Lieb}, \citenamefont {Schultz},\ and\ \citenamefont {Mattis}}]{lieb1961two}%
  \BibitemOpen
  \bibfield  {author} {\bibinfo {author} {\bibfnamefont {E.}~\bibnamefont {Lieb}}, \bibinfo {author} {\bibfnamefont {T.}~\bibnamefont {Schultz}},\ and\ \bibinfo {author} {\bibfnamefont {D.}~\bibnamefont {Mattis}},\ }\bibfield  {title} {\bibinfo {title} {Two soluble models of an antiferromagnetic chain},\ }\href@noop {} {\bibfield  {journal} {\bibinfo  {journal} {Annals of Physics}\ }\textbf {\bibinfo {volume} {16}},\ \bibinfo {pages} {407} (\bibinfo {year} {1961})}\BibitemShut {NoStop}%
\bibitem [{\citenamefont {Mazziotti}(2006)}]{mazziotti2006variational}%
  \BibitemOpen
  \bibfield  {author} {\bibinfo {author} {\bibfnamefont {D.~A.}\ \bibnamefont {Mazziotti}},\ }\bibfield  {title} {\bibinfo {title} {Variational reduced-density-matrix method using three-particle n-representability conditions with application to many-electron molecules},\ }\href@noop {} {\bibfield  {journal} {\bibinfo  {journal} {Physical Review A}\ }\textbf {\bibinfo {volume} {74}},\ \bibinfo {pages} {032501} (\bibinfo {year} {2006})}\BibitemShut {NoStop}%
\bibitem [{\citenamefont {Mazziotti}(2004)}]{mazziotti2004realization}%
  \BibitemOpen
  \bibfield  {author} {\bibinfo {author} {\bibfnamefont {D.~A.}\ \bibnamefont {Mazziotti}},\ }\bibfield  {title} {\bibinfo {title} {Realization of quantum chemistry without wave functions through first-order semidefinite programming},\ }\href@noop {} {\bibfield  {journal} {\bibinfo  {journal} {Physical review letters}\ }\textbf {\bibinfo {volume} {93}},\ \bibinfo {pages} {213001} (\bibinfo {year} {2004})}\BibitemShut {NoStop}%
\bibitem [{\citenamefont {Navascu{\'e}s}\ \emph {et~al.}(2007)\citenamefont {Navascu{\'e}s}, \citenamefont {Pironio},\ and\ \citenamefont {Ac{\'\i}n}}]{navascues2007bounding}%
  \BibitemOpen
  \bibfield  {author} {\bibinfo {author} {\bibfnamefont {M.}~\bibnamefont {Navascu{\'e}s}}, \bibinfo {author} {\bibfnamefont {S.}~\bibnamefont {Pironio}},\ and\ \bibinfo {author} {\bibfnamefont {A.}~\bibnamefont {Ac{\'\i}n}},\ }\bibfield  {title} {\bibinfo {title} {Bounding the set of quantum correlations},\ }\href@noop {} {\bibfield  {journal} {\bibinfo  {journal} {Physical Review Letters}\ }\textbf {\bibinfo {volume} {98}},\ \bibinfo {pages} {010401} (\bibinfo {year} {2007})}\BibitemShut {NoStop}%
\bibitem [{\citenamefont {Rai}\ \emph {et~al.}(2024)\citenamefont {Rai}, \citenamefont {Kull}, \citenamefont {Emonts}, \citenamefont {Tura}, \citenamefont {Schuch},\ and\ \citenamefont {Baccari}}]{rai2024hierarchy}%
  \BibitemOpen
  \bibfield  {author} {\bibinfo {author} {\bibfnamefont {K.~S.}\ \bibnamefont {Rai}}, \bibinfo {author} {\bibfnamefont {I.}~\bibnamefont {Kull}}, \bibinfo {author} {\bibfnamefont {P.}~\bibnamefont {Emonts}}, \bibinfo {author} {\bibfnamefont {J.}~\bibnamefont {Tura}}, \bibinfo {author} {\bibfnamefont {N.}~\bibnamefont {Schuch}},\ and\ \bibinfo {author} {\bibfnamefont {F.}~\bibnamefont {Baccari}},\ }\bibfield  {title} {\bibinfo {title} {A hierarchy of spectral gap certificates for frustration-free spin systems},\ }\href@noop {} {\bibfield  {journal} {\bibinfo  {journal} {arXiv preprint arXiv:2411.03680}\ } (\bibinfo {year} {2024})}\BibitemShut {NoStop}%
\bibitem [{\citenamefont {Javadi-Abhari}\ \emph {et~al.}(2024)\citenamefont {Javadi-Abhari}, \citenamefont {Treinish}, \citenamefont {Krsulich}, \citenamefont {Wood}, \citenamefont {Lishman}, \citenamefont {Gacon}, \citenamefont {Martiel}, \citenamefont {Nation}, \citenamefont {Bishop}, \citenamefont {Cross} \emph {et~al.}}]{javadi2024quantum}%
  \BibitemOpen
  \bibfield  {author} {\bibinfo {author} {\bibfnamefont {A.}~\bibnamefont {Javadi-Abhari}}, \bibinfo {author} {\bibfnamefont {M.}~\bibnamefont {Treinish}}, \bibinfo {author} {\bibfnamefont {K.}~\bibnamefont {Krsulich}}, \bibinfo {author} {\bibfnamefont {C.~J.}\ \bibnamefont {Wood}}, \bibinfo {author} {\bibfnamefont {J.}~\bibnamefont {Lishman}}, \bibinfo {author} {\bibfnamefont {J.}~\bibnamefont {Gacon}}, \bibinfo {author} {\bibfnamefont {S.}~\bibnamefont {Martiel}}, \bibinfo {author} {\bibfnamefont {P.~D.}\ \bibnamefont {Nation}}, \bibinfo {author} {\bibfnamefont {L.~S.}\ \bibnamefont {Bishop}}, \bibinfo {author} {\bibfnamefont {A.~W.}\ \bibnamefont {Cross}}, \emph {et~al.},\ }\bibfield  {title} {\bibinfo {title} {Quantum computing with qiskit},\ }\href@noop {} {\bibfield  {journal} {\bibinfo  {journal} {arXiv preprint arXiv:2405.08810}\ } (\bibinfo {year} {2024})}\BibitemShut {NoStop}%
\bibitem [{\citenamefont {Agrawal}\ \emph {et~al.}(2018)\citenamefont {Agrawal}, \citenamefont {Verschueren}, \citenamefont {Diamond},\ and\ \citenamefont {Boyd}}]{agrawal2018rewriting}%
  \BibitemOpen
  \bibfield  {author} {\bibinfo {author} {\bibfnamefont {A.}~\bibnamefont {Agrawal}}, \bibinfo {author} {\bibfnamefont {R.}~\bibnamefont {Verschueren}}, \bibinfo {author} {\bibfnamefont {S.}~\bibnamefont {Diamond}},\ and\ \bibinfo {author} {\bibfnamefont {S.}~\bibnamefont {Boyd}},\ }\bibfield  {title} {\bibinfo {title} {A rewriting system for convex optimization problems},\ }\href@noop {} {\bibfield  {journal} {\bibinfo  {journal} {Journal of Control and Decision}\ }\textbf {\bibinfo {volume} {5}},\ \bibinfo {pages} {42} (\bibinfo {year} {2018})}\BibitemShut {NoStop}%
\bibitem [{\citenamefont {Diamond}\ and\ \citenamefont {Boyd}(2016)}]{diamond2016cvxpy}%
  \BibitemOpen
  \bibfield  {author} {\bibinfo {author} {\bibfnamefont {S.}~\bibnamefont {Diamond}}\ and\ \bibinfo {author} {\bibfnamefont {S.}~\bibnamefont {Boyd}},\ }\bibfield  {title} {\bibinfo {title} {{CVXPY}: {A} {P}ython-embedded modeling language for convex optimization},\ }\href@noop {} {\bibfield  {journal} {\bibinfo  {journal} {Journal of Machine Learning Research}\ }\textbf {\bibinfo {volume} {17}},\ \bibinfo {pages} {1} (\bibinfo {year} {2016})}\BibitemShut {NoStop}%
\bibitem [{\citenamefont {Markovich}\ \emph {et~al.}(2024)\citenamefont {Markovich}, \citenamefont {Malikis}, \citenamefont {Polla},\ and\ \citenamefont {Tura}}]{markovich2024parameter}%
  \BibitemOpen
  \bibfield  {author} {\bibinfo {author} {\bibfnamefont {L.}~\bibnamefont {Markovich}}, \bibinfo {author} {\bibfnamefont {S.}~\bibnamefont {Malikis}}, \bibinfo {author} {\bibfnamefont {S.}~\bibnamefont {Polla}},\ and\ \bibinfo {author} {\bibfnamefont {J.}~\bibnamefont {Tura}},\ }\bibfield  {title} {\bibinfo {title} {Parameter shift rule with optimal phase selection},\ }\href@noop {} {\bibfield  {journal} {\bibinfo  {journal} {Physical Review A}\ }\textbf {\bibinfo {volume} {109}},\ \bibinfo {pages} {062429} (\bibinfo {year} {2024})}\BibitemShut {NoStop}%
\bibitem [{\citenamefont {Schuch}\ and\ \citenamefont {Cirac}(2010)}]{schuch2010matrix}%
  \BibitemOpen
  \bibfield  {author} {\bibinfo {author} {\bibfnamefont {N.}~\bibnamefont {Schuch}}\ and\ \bibinfo {author} {\bibfnamefont {J.~I.}\ \bibnamefont {Cirac}},\ }\bibfield  {title} {\bibinfo {title} {Matrix product state and mean-field solutions for one-dimensional systems can be found efficiently},\ }\href@noop {} {\bibfield  {journal} {\bibinfo  {journal} {Physical Review A—Atomic, Molecular, and Optical Physics}\ }\textbf {\bibinfo {volume} {82}},\ \bibinfo {pages} {012314} (\bibinfo {year} {2010})}\BibitemShut {NoStop}%
\bibitem [{\citenamefont {Anselmetti}\ \emph {et~al.}(2024)\citenamefont {Anselmetti}, \citenamefont {Degroote}, \citenamefont {Moll}, \citenamefont {Santagati},\ and\ \citenamefont {Streif}}]{anselmetti2024classical}%
  \BibitemOpen
  \bibfield  {author} {\bibinfo {author} {\bibfnamefont {G.-L.~R.}\ \bibnamefont {Anselmetti}}, \bibinfo {author} {\bibfnamefont {M.}~\bibnamefont {Degroote}}, \bibinfo {author} {\bibfnamefont {N.}~\bibnamefont {Moll}}, \bibinfo {author} {\bibfnamefont {R.}~\bibnamefont {Santagati}},\ and\ \bibinfo {author} {\bibfnamefont {M.}~\bibnamefont {Streif}},\ }\bibfield  {title} {\bibinfo {title} {Classical optimisation of reduced density matrix estimations with classical shadows using n-representability conditions under shot noise considerations},\ }\href@noop {} {\bibfield  {journal} {\bibinfo  {journal} {arXiv preprint arXiv:2411.18430}\ } (\bibinfo {year} {2024})}\BibitemShut {NoStop}%
\bibitem [{\citenamefont {Avdic}\ and\ \citenamefont {Mazziotti}(2024)}]{avdic2024fewer}%
  \BibitemOpen
  \bibfield  {author} {\bibinfo {author} {\bibfnamefont {I.}~\bibnamefont {Avdic}}\ and\ \bibinfo {author} {\bibfnamefont {D.~A.}\ \bibnamefont {Mazziotti}},\ }\bibfield  {title} {\bibinfo {title} {Fewer measurements from shadow tomography with n-representability conditions},\ }\href@noop {} {\bibfield  {journal} {\bibinfo  {journal} {Physical Review Letters}\ }\textbf {\bibinfo {volume} {132}},\ \bibinfo {pages} {220802} (\bibinfo {year} {2024})}\BibitemShut {NoStop}%
\bibitem [{\citenamefont {Rosset}\ \emph {et~al.}(2012)\citenamefont {Rosset}, \citenamefont {Ferretti-Sch\"obitz}, \citenamefont {Bancal}, \citenamefont {Gisin},\ and\ \citenamefont {Liang}}]{PhysRevA.86.062325}%
  \BibitemOpen
  \bibfield  {author} {\bibinfo {author} {\bibfnamefont {D.}~\bibnamefont {Rosset}}, \bibinfo {author} {\bibfnamefont {R.}~\bibnamefont {Ferretti-Sch\"obitz}}, \bibinfo {author} {\bibfnamefont {J.-D.}\ \bibnamefont {Bancal}}, \bibinfo {author} {\bibfnamefont {N.}~\bibnamefont {Gisin}},\ and\ \bibinfo {author} {\bibfnamefont {Y.-C.}\ \bibnamefont {Liang}},\ }\bibfield  {title} {\bibinfo {title} {Imperfect measurement settings: Implications for quantum state tomography and entanglement witnesses},\ }\href {https://doi.org/10.1103/PhysRevA.86.062325} {\bibfield  {journal} {\bibinfo  {journal} {Phys. Rev. A}\ }\textbf {\bibinfo {volume} {86}},\ \bibinfo {pages} {062325} (\bibinfo {year} {2012})}\BibitemShut {NoStop}%
\end{thebibliography}%

\end{document}